# RESTITUTION OF THE TEMPERATURE FIELD INSIDE A CYLINDER OF SEMI-TRANSPARENT DENSE MEDIUM FROM DIRECTIONAL INTENSITY DATA

V. LE DEZ, D. LEMONNIER and H. SADAT

Laboratoire d'Etudes Thermiques UMR 6608 CNRS-ENSMA - 86960 Futuroscope Cedex, France

**Abstract** - The purpose of this paper is to obtain the temperature field inside a cylinder filled in with a dense non scattering semi-transparent medium from directional intensity data by solving the inverse radiative transfer equation. This equation is solved in a first approach with the help of a discrete scheme and the solution is then exactly obtained by separating the physical set on two disjoint domains on which a Laplace transform is applied, followed by the resolution of a 1st kind Fredholm equation.

**NOMENCLATURE**:

| | |
|---|---|
| $R$ | *radius of the cylinder (m)* |
| $\kappa_\lambda$ | *spectral absorption coefficient* ($m^{-1}$) |
| $n_\lambda$ | *spectral refractive index* |
| $T$ | *temperature (K)* |
| $L_\lambda(x)$ | *directional monochromatic intensity* ($W.m^{-3}.Sr^{-1}$) |
| $L_\lambda^0(T)$ | *Planck function (black body intensity)* |
| $g(x)$ | *data set* |
| $\bar{\rho}(x),\ \rho_\perp(x),\ \rho_{\|\|}(x)$ | *mean reflection factor, perpendicular and parallel polarisation reflection factors* |
| $h(x)$ | *generalized data set* |
| $\psi(x)$ | *data function on the range* $x > \frac{R}{n_\lambda}$ |
| $x_i$ | *centre of a discrete cell labelled i* |
| $N$ | *number of discrete cells in the cylinder* |
| $\Delta r$ | *depth of a cell (m)* |
| $E(x)$ | *integer part of a real number x* |
| $Ci$, $Chi$ | *cosine integral, hyperbolic cosine integral* |
| $Ei$, $E_1$ | *exponential integral, exponential integral of 1st order* |
| $\alpha$ | *regularization parameter* |
| $\delta$ | *noise intensity factor* |
| $\tau = \kappa_\lambda r$, $\tau_0 = \kappa_\lambda R$ | *optical depths* |
| $\gamma$ | *Euler-Mascheroni constant* |
| $K(w,z')$ | *non symmetric kernel of integral equation* |
| $K_G$, $K_D$ | *symmetric left and right kernels* |
| $(\lambda_n)_{n\in\|N}$, $(\mu_n)_{n\in\|N}$ | *eigenvalues and eigenfunctions of the right kernel* $K_D$ |

## 1. INTRODUCTION

The determination of a temperature field inside an axisymmetric semi-transparent medium from optical infrared measurements has been studied many decades ago, especially for gaseous media where the refractive index can be ignored, in the sense where it remains close to one for a large scale of wavelengths. Indeed, the energetic spectral and/or directional radiative flux emerging from such a medium allows the reconstruction of the internal thermal distribution generating this flux, and from this basic principle Milne [1] was a pioneer in retrieving the temperature field in the superficial area of the sun from directional and spectral intensities, followed by Chahine [2] who determined atmospheric temperature profiles from spectral outgoing radiances. A similar calculation was performed by Ben Abdallah [3] who studied the gaseous atmosphere of a giant planet from spectral intensities, including an elegant regularisation method to take into account the presence of noisy data: in this latter case, a linearization of the Planck function was done to separate the spectral and temperature dependencies. Siewert [4] extended an equivalent approach with the help of orthogonal functions development to the determination of the internal source in an absorbing and scattering sphere from directional emerging intensity data, and Li [5] achieved the restitution of the temperature field inside a cylindrical medium from emerging intensities with a standard functional minimization, ignoring the transmission refractive effects generally induced with specularly transparent reflecting surfaces, due here to a constant reflection factor. As mentioned before, in the most studies dealing with gaseous atmospheres, the refractive index of the medium is generally taken as one, so that no reflective and refractive effect can affect the transmitted intensities. In this latter situation, an exact solution can commonly be obtained with the help of a Laplace transform when no scattering occurs. A significant improvement was brought by Kocifaj [6] who took into account the long distance ray deviation due to the continuous refractive effects by solving simultaneously two inverse problems. More specifically devoted to the cylindrical geometry, the study of Liu et al. [7] allows similarly the reconstruction of absorption and temperature profiles inside gaseous axisymmetric flames. Considering two distinct but analogous problems, they reconstruct the absorption field by using an Abel equation and perform a standard minimisation for the temperature profile, due to the non constant absorption, and note that a conjugate gradient method taking into account the sensibility matrix gives in this special case accurate results when treating noisy data without any regularisation technique.

Nevertheless, for dense media such as glasses, of significantly higher absorption coefficients, the refractive index is much greater than one and such effects can no longer be ignored. Although several papers have been devoted to the restitution of the inner temperature field, or the radiative source field, for plane dense media parallel slabs so as for Cartesian and cylindrical geometry devices, [8-10] for instance who apply an objective functional minimisation mainly based on the conjugate gradient technique taking into account the sensibility matrix and looking at the noisy data influence on the numerical procedure, very little literature to our knowledge was interested in retrieving the temperature field inside cylindrical semi-transparent non gaseous media with specularly reflecting surfaces, from emerging intensity measurements.

Viskanta et al. [11] used restitution techniques from deep area optical measurements, but due to the particular geometry of the system, only spectral intensities where investigated. Generally devoted to the "best" way in regularising the experimental data for determining internal sources from emerging intensities, like in [12, 13], these approaches for dense media do not take advantage of the particular properties of the operator governing the problem's physics, specially interesting in cylindrical geometry. Ertürk et al. [10] applied three different regularisation techniques when retrieving an internal radiative source field inside three-dimensional device, or inverse boundary condition estimation. Since the discrete associated problem often reduces to a linear system of equations resulting from a Fredholm equation of the first kind, they deduce that the corresponding problem is ill-posed and need be regularized to avoid error amplifications. They use a truncated singular value decomposition (TSVD) and two related conjugate gradient methods which produce analogous and relatively accurate results for their problems, and conclude that a simple TSVD of easy computation is generally an efficient technique.

In the case of axisymmetric systems, most of the authors recommend to use both spectral and directional measurements: Natterer [14] and Sakami et al. [15] indicate that directional measurements do not allow the complete restitution of the internal temperature field because of refractive effects, and use spectral measurements to perform the complete reconstruction in a so-called “missing data problem”. In spite of this, they indicate that such a technique may be limited because of the absorption spectrum sensibility, since dense materials have smooth absorption coefficient spectral fields in the transparency band. This major problem, i.e. a weak sensibility with respect to the absorption for dense media, can only give approximate temperature fields on a few numbers of points, including a priori information on the field. That is why one should prefer directional monochromatic intensities when reconstructing a complete thermal field inside a dense cylinder.

In this paper, we determine the complete temperature field expression inside a cylinder of radius $R$ filled with a dense semi-transparent medium of non unit refractive index $n_\lambda$. We first solve the associated inverse problem by using a numerical approach. It is shown that a simple LU decomposition of the associated matrix gives accurate results only in a restricted area $x \le \frac{R}{n_\lambda}$ and fails in the complement region $x > \frac{R}{n_\lambda}$, even for perfect theoretical non noisy outgoing intensities. This later observation crudely differs from what happens in media of unit refractive index such as gaseous atmospheres, where a simple LU decomposition gives exact results everywhere in the cylinder for non noisy data, and no regularization need be applied on the input data. When the refractive index is 1, a simple TSVD [9] for instance, gives accurate results also when adding perturbations in the input data set. Here, for refractive indices strictly greater than 1, a TSVD must be performed even with perfect non noisy input data to obtain acceptable results in the region $x > \frac{R}{n_\lambda}$, and only for relatively moderate refractive indices. The results are even accurate when the number of discrete nodes is lower than a threshold value (otherwise the method fails), strongly depending on the medium’s refractive index. The scope of this article is then to understand why a regularization has to be performed on the global problem even in presence of perfect input data, contrarily to what happens for unit refractive indices, and will avoid the specific problem of regularizing the input data. To do so, the input data are considered as perfect non noisy data obtained from a direct calculation. The exact solution of the integral equation governing the related temperature field is proposed by separating the whole problem into two disjoint calculations on two separate sets. This allows us to finally show that the problem does not belong to the “missing data problems”, as frequently mentioned, although a discussion on the eigenvalues of the operator’s kernel demonstrate why this solution is impossible to obtain in a practical way.

## 2. GEOMETRICAL AND PHYSICAL MODEL

One considers a cylinder of radius *R* filled with dense absorbing-emitting but non scattering semi-transparent medium (STM), such as glass or crystal, mainly characterised by its spectral absorption coefficient $\kappa_\lambda$ and its spectral refractive index $n_\lambda$. The medium (amorphous or not) being supposed made of dense bulk material without heterogeneity, is weakly or not scattering, and its absorption and refraction fields do not strongly depend on the inner temperature for relatively large temperature scales. Furthermore, the absorption (and refraction) spectrum of dense media being extremely smooth on the major part of the semi-transparency band, spectral measurements of the emerging intensities emitted by the body will present an extremely bad sensitivity relatively to the wavelength, and angular collected monochromatic (simulated) data have been preferred to a spectral intensities set, at a wavelength where the lateral surface of the cylinder is not opaque.

The cylinder is submitted to a given spatial temperature field (function of the radial position *r* and the longitudinal one *z* along the cylinder’s axis), and emits an infrared radiation which is collected and converted into emerging intensities, function of the spatial position, and also depending on the absorption coefficient, the refractive index and the temperature field. The directional intensities are collected in a tomographic plane perpendicular to the cylinder’s axis at a given *z*, which allows the problem to be considered as one-dimensional, i.e. the emerging intensities at a given *z* are only depending on the geometric variable *x* due to the revolution symmetry.

Since the refractive index is assumed to be strictly greater than one, the reflection coefficients at the lateral specular surface between the semi-transparent medium and the surrounding environment, given by the Fresnel formulae, are function of the local position *x*. The geometrical path in the tomographic plane for a particular emerging ray consists in a complete series of broken lines as illustrated on Fig. 1a

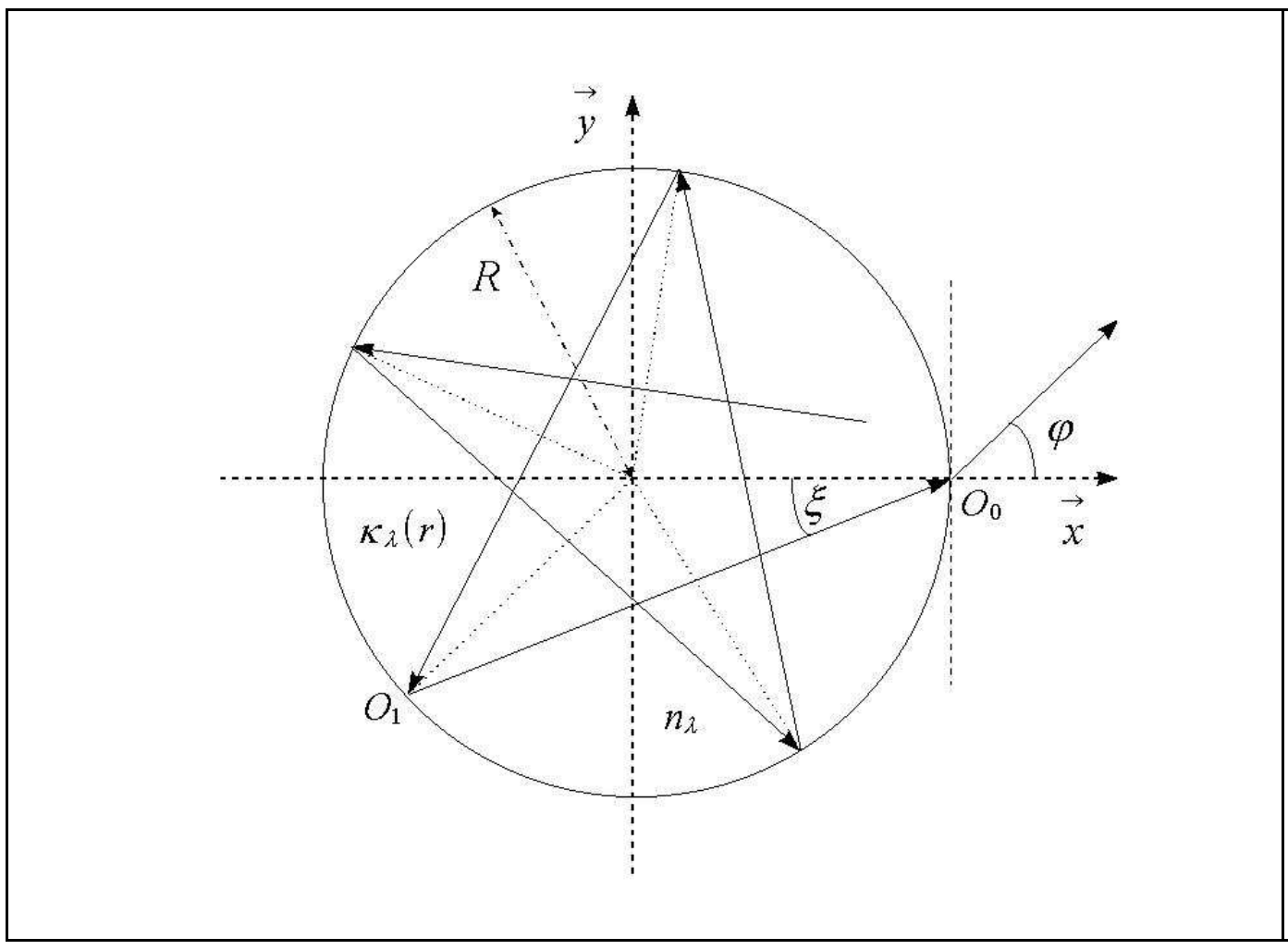

Fig. 1a: internal trajectory of a ray emerging in the tomographic plane

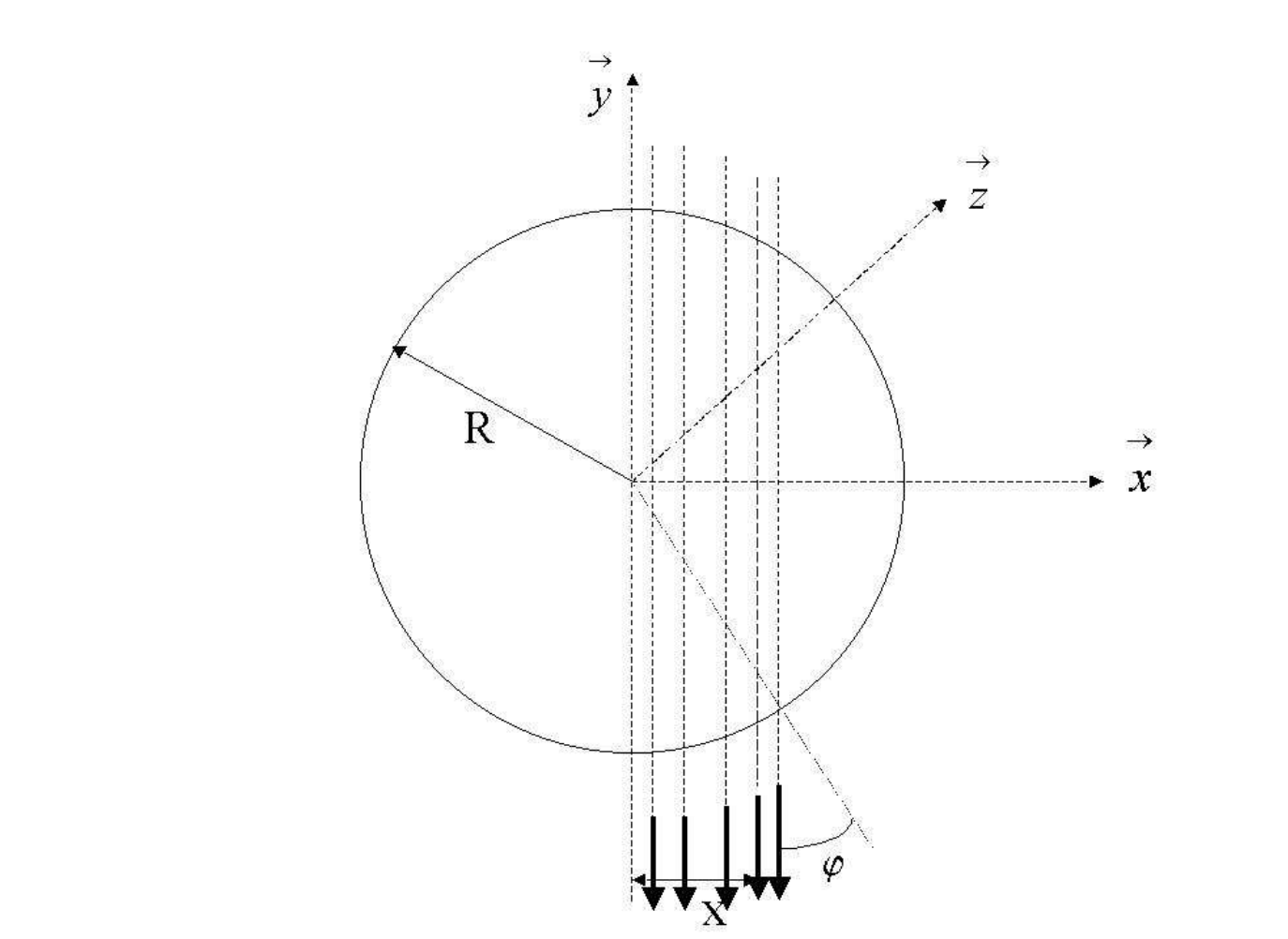

Fig. 1b: schematic description of the emerging intensities at local position $0 \le x \le R$

The outgoing intensity $L_\lambda(x)$ at position *x* is given by the classical expression [16]

$$L_\lambda(x) = \frac{2\,\kappa_\lambda\,[1-\rho(x)]\,e^{-\kappa_\lambda R\cos\xi}}{1-\rho(x)\,e^{-2\kappa_\lambda R\cos\xi}} \int_{r=R\sin\xi}^{R} \frac{r\,L^0_\lambda[T(r)]}{\sqrt{r^2-R^2\sin^2\xi}}\,\cosh\left(\kappa_\lambda\sqrt{r^2-R^2\sin^2\xi}\right)dr \qquad (1)$$

In this expression, $L_\lambda(x)$ is the directional monochromatic intensity emerging from the cylinder in a plane $\left(\vec{x},\vec{y}\right)$ orthogonal to the cylinder's axis $\vec{z}$, for a given angle $\varphi$ (between the emerging ray at abscissa *x* and the normal to the lateral surface of the cylinder at this point), simply related to the spatial position *x* by $x = R\sin\varphi$, as depicted on the schematic figure 1b. The spectral absorption coefficient is assumed to be known and spatially constant, and the reflection factors $\rho(x)$, either parallel or perpendicular, are classically given by the Fresnel formulae for transparent media, considering that the imaginary part of the complex refractive index is much lower than the real one in the semi-transparent band.

$T(r)$ is the spatial temperature field to be reconstructed and $L^0_\lambda$ the Planck function. Noticing that from Descartes' law $\sin\varphi = n_\lambda \sin\xi$, it obviously comes $x = n_\lambda R\sin\xi$, and Eq. (1) is equivalent to the following integral equation

$$\int_{r=\frac{x}{n_\lambda}}^{R} \frac{r\,\kappa_\lambda\,L^0_\lambda[T(r)]}{\sqrt{r^2-\frac{x^2}{n_\lambda^2}}}\,\cosh\left(\kappa_\lambda\sqrt{r^2-\frac{x^2}{n_\lambda^2}}\right)dr = \frac{\left[1-\bar{\rho}(x)\,e^{-2\kappa_\lambda\sqrt{R^2-\frac{x^2}{n_\lambda^2}}}\right]L_\lambda(x)}{2\left[1-\bar{\rho}(x)\right]e^{-\kappa_\lambda\sqrt{R^2-\frac{x^2}{n_\lambda^2}}}} = g(x) \qquad (2)$$

The unknown function is $L^0_\lambda[T(r)]$, the useful data being the discrete set $g(x)$, where the mean reflection factor is defined by $\bar{\rho}(x) = \frac{1}{2}\left[\rho_\perp(x)+\rho_{||}(x)\right]$. In the previous expression, the subscript $\perp$ stands for the reflection factor related to the perpendicular polarisation, while the subscript $||$ denotes the parallel polarisation.

The behaviour of the directional emerging intensity with respect to the local position *x* inside the cylinder has been reported on Figs 2a-b, when the internal temperature field inside the cylinder is assumed linear, with $T(x) = 273.15 + 200.\left(2 - \frac{x}{R}\right)$, for $n_\lambda = 1.5$ and $n_\lambda = 4.5$. It appears on these results that increasing the refractive index apparently acts on the emerging intensities as increasing the absorption coefficient for a fixed lower refractive index.

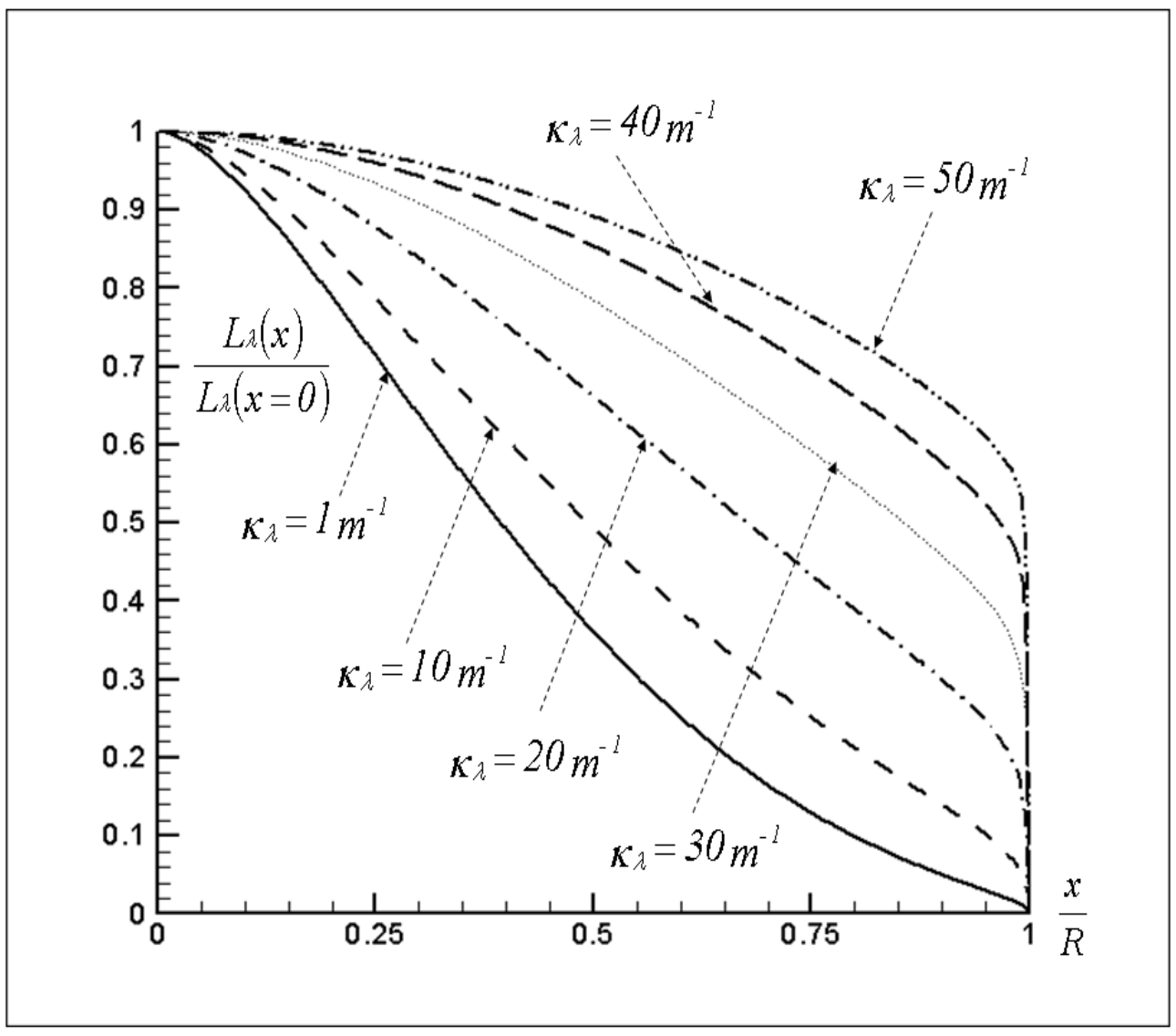

Fig 2a: intensity for a linear temperature field, $n_\lambda = 1.5$

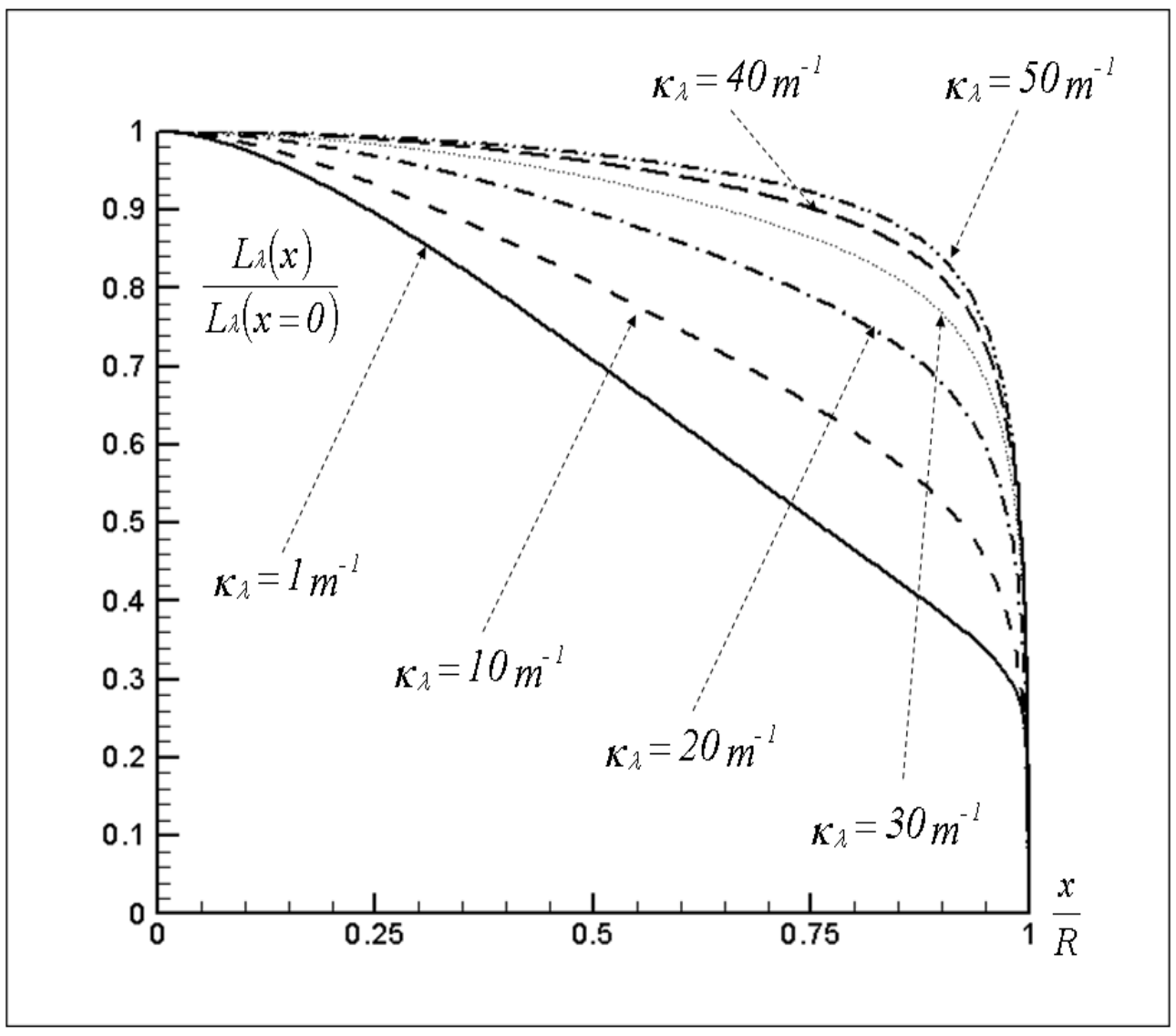

Fig 2b: intensity for a linear temperature field, $n_\lambda = 4.5$

The physical model being fully described, the following section shall be now devoted to the numerical discretization of Eq. (2) to obtain the temperature field $T(r)$.

## 3. DISCRETE NUMERICAL SOLUTION OF THE INTEGRAL EQUATION

### 3.1 Numerical Scheme

The continuous Eq. (2) is transformed into a linear system of *N* equations by using a spatial discretization. One defines *N* control volumes (cells) of depth $\Delta r$, labelled *i* and whose centre is characterised by $x_i = \frac{i-1}{N-1} R = (i-1)\Delta r$ for $1 \le i \le N$. In each cell, the temperature is supposed constant, with $T = T_1$ on $\left[x_1 = 0, \frac{\Delta r}{2}\right[$, $T = T_i$ on $\left[x_i - \frac{\Delta r}{2}, x_i + \frac{\Delta r}{2}\right[$ for $2 \le i \le N-1$, and $T = T_N$ on $\left[R - \frac{\Delta r}{2}, x_N = R\right]$.

For a refractive index strictly greater than 1, $\frac{x_i}{n_\lambda} < x_i$, and it exists an integer $p$ strictly lower than $i$, i.e. $p \in [1, i-1]$, such that $x_p \leq \frac{x_i}{n_\lambda} < x_{p+1}$. This simply means that for a refractive index very close to 1, $\frac{x_i}{n_\lambda} \approx x_i$ and this point belongs to the cell which contains $x_i$, i.e. $x_{i-1} \leq \frac{x_i}{n_\lambda} < x_i$, and $p = i-1$. On the other hand, for large refractive indices, the point characterized by its position $\frac{x_i}{n_\lambda}$ is much closer from the cylinder's centre than $x_i$ and belongs to a cell far from the one containing $x_i$. As an example, let us choose a cylinder of radius $R = 24$ cm and $n_\lambda = 6$, with $N = 25$: for $i = 13$, $x_i = 12$ cm while $\frac{x_i}{n_\lambda} = 2$ cm, so that $\frac{x_i}{n_\lambda} = x_3$ and $p = 3$. .

A simple analysis shows that the corresponding integer $p$ is defined by $p = 1 + E\left(\frac{i-1}{n_\lambda}\right)$ where $E$ stands for the integer part of a real number $x$, with $E(x) = m$ and $m \leq x < m+1$. Hence defining the index $q$ by

$$q = \begin{cases} p & \text{if } x_p \leq \frac{x_i}{n_\lambda} < x_p + \frac{\Delta r}{2} \\ p+1 & \text{if } x_p + \frac{\Delta r}{2} \leq \frac{x_i}{n_\lambda} < x_{p+1} \end{cases}$$

, the discrete form of Eq. (2) can be formulated by the following system for $1 \leq i \leq N-1$:

$$\begin{aligned} g(x_i) = {} & L_\lambda^0(T_q) \sinh\left[\kappa_\lambda \sqrt{\left(x_q + \frac{\Delta r}{2}\right)^2 - \frac{x_i^2}{n_\lambda^2}}\right] \\ & + \sum_{k=q+1}^{N-1} L_\lambda^0(T_k) \left\{ \sinh\left[\kappa_\lambda \sqrt{\left(x_k + \frac{\Delta r}{2}\right)^2 - \frac{x_i^2}{n_\lambda^2}}\right] - \sinh\left[\kappa_\lambda \sqrt{\left(x_k - \frac{\Delta r}{2}\right)^2 - \frac{x_i^2}{n_\lambda^2}}\right] \right\} \\ & + L_\lambda^0(T_N) \left\{ \sinh\left(\kappa_\lambda \sqrt{R^2 - \frac{x_i^2}{n_\lambda^2}}\right) - \sinh\left[\kappa_\lambda \sqrt{\left(R - \frac{\Delta r}{2}\right)^2 - \frac{x_i^2}{n_\lambda^2}}\right] \right\} \end{aligned} \qquad (3)$$

The case $i = N$ is a particular case, since, although $L_\lambda(x_N) = L_\lambda(R) = 0$ is known, $g(x_N)$ is non equal to 0 (except for a unit refractive index) and undetermined because $\rho(x_N) = \rho(R) = 1$: then $g(x_N)$ has to be extrapolated, for instance from the other values.

Defining the index $t$ as $t=\begin{cases} j_{max} & if\ x_{j_{max}} \le \frac{R}{n_\lambda} < x_{j_{max}} + \frac{\Delta r}{2} \\ j_{max}+1 & if\ x_{j_{max}} + \frac{\Delta r}{2} \le \frac{R}{n_\lambda} < x_{j_{max}+1} \end{cases}$, where $j_{max}=1+E\left(\frac{N-1}{n_\lambda}\right)$, the discrete form of Eq. (2) for $i=N$ is identical to Eq. (4) if $t \le N-1$, while if $t = N$, it reduces to:

$$g(x_N)=L^0_\lambda(T_N)\,sinh\left(\kappa_\lambda R\sqrt{1-\frac{1}{n_\lambda^2}}\right) \tag{4}$$

Hence Eqs. (3-4) can be transformed into the following linear system whose solution gives the unknown intensities:

$$C\,L^0{}_\lambda\ (T)=g \tag{5}$$

**3.2 Numerical results**

In all what follows, the numerical applications will be done for a cylinder of radius $R$ = 24 cm, filled with a semi-transparent medium characterised by its refractive index $n_\lambda=1.5$ or $n_\lambda=4.5$ and its spectral absorption coefficient $\kappa_\lambda=10\ m^{-1}$ at wavelength $\lambda=1.5\ \mu m$, the temperature field being given by $T(x)=573.15+100.\left(1-e^{-\frac{5x}{2R}}\right)sin\left(\frac{11\pi x}{2R}\right)$.

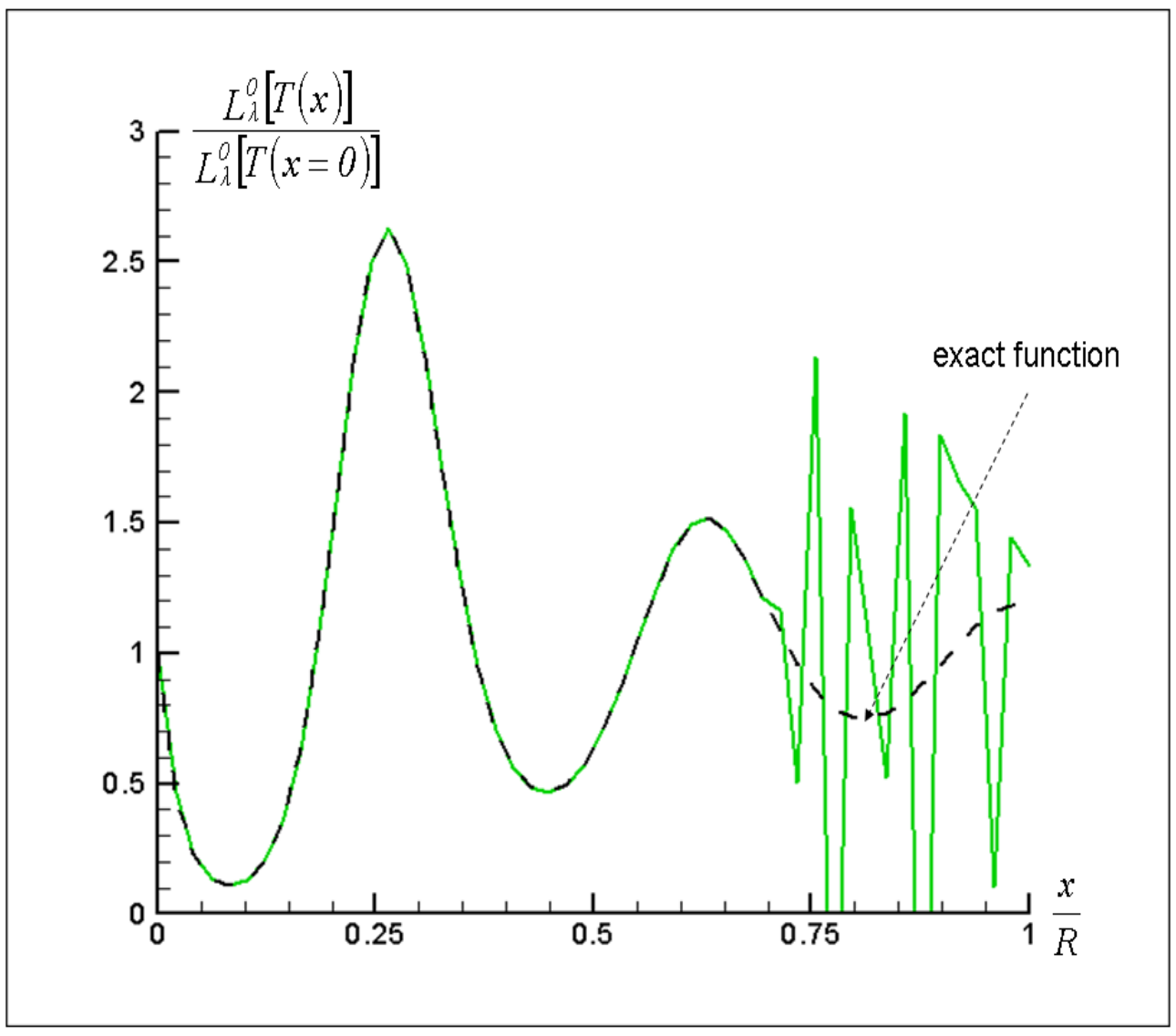

Fig 3a: retrieved Planck function for $N=50$

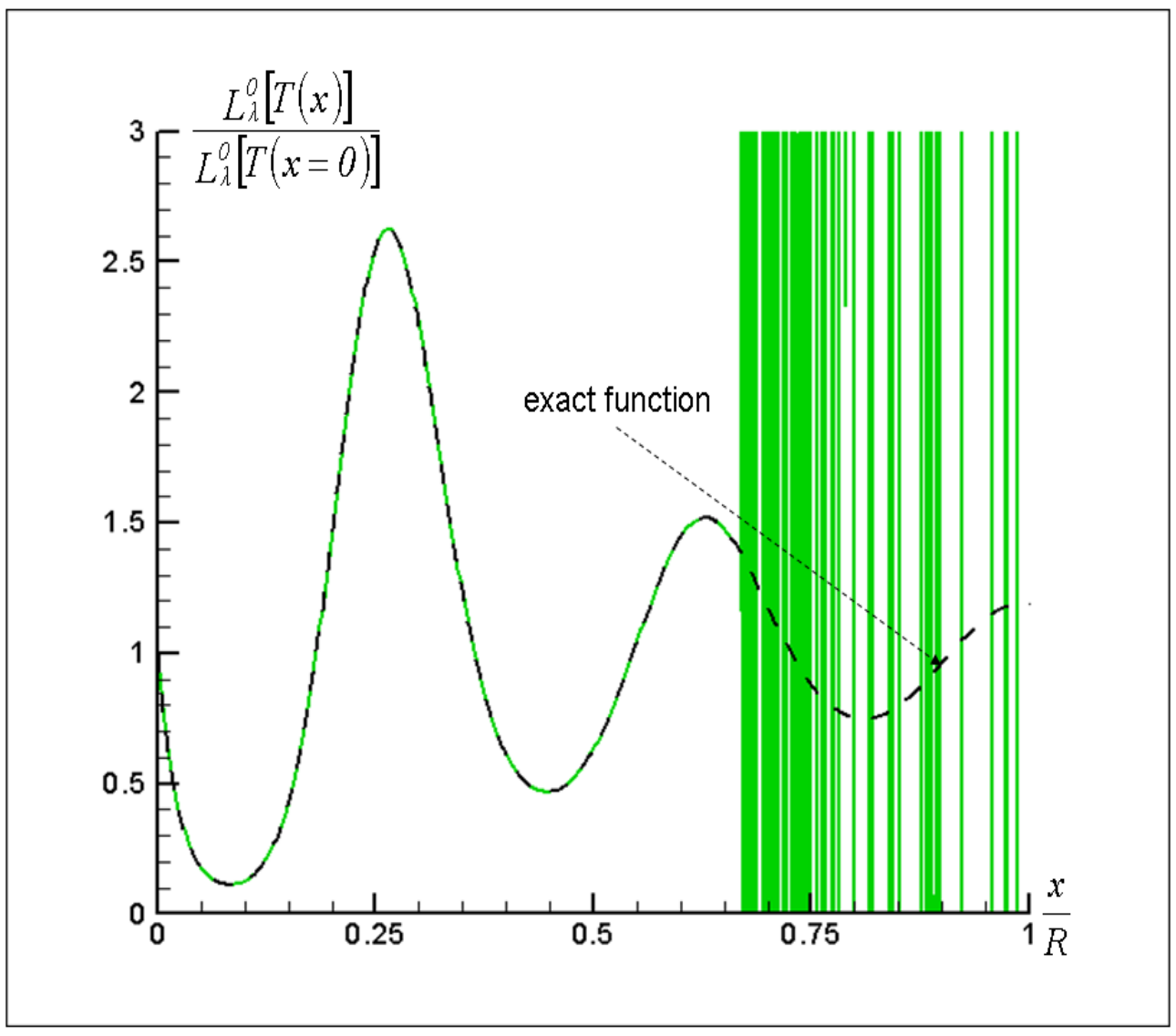

Fig 3b: retrieved Planck function for $N=500$

The numerical results obtained by solving to the former linear system are presented on Figs 3a, b for $N=50$ and $500$ cells. A simple direct LU decomposition of matrix $C$ has been used. It can be seen that the Planck function is correctly determined for $x \le \frac{R}{n_\lambda}$. The method fails however in the area $x > \frac{R}{n_\lambda}$, even with exact non noisy data. The results are oscillatory and inaccurate with an error increasing rapidly with the grid refinement. However, the complete temperature field is correctly evaluated for $N$ lower than 27 in this particular situation. For $N \le 26$ and perfect data, the retrieved Planck function field cannot be distinguished from the exact Planck function at all internal point inside the medium, while as soon as $N \ge 27$, the retrieved field is of poor quality in the neighbourhood of the lateral surface of the cylinder.
This threshold value of the number of cells, below which the reconstructed field is correctly determined for perfect data, is strongly depending on the refractive index of the medium, as clearly shown on figures 4a and 4b. In these examples, the number of cells is set to $N = 27$. For $n_\lambda = 1.5$, the number of cells equals the threshold value and the reconstructed field is extremely close to the exact one, except in a small neighbourhood of the lateral surface. On the contrary, when $n_\lambda = 4.5$, the reconstructed field is correctly estimated only on the range $x \le \frac{R}{n_\lambda}$ and of extremely bad quality for $x > \frac{R}{n_\lambda}$. In this case the threshold value is $N = 9$ and lower than the number of cells used in the calculation.

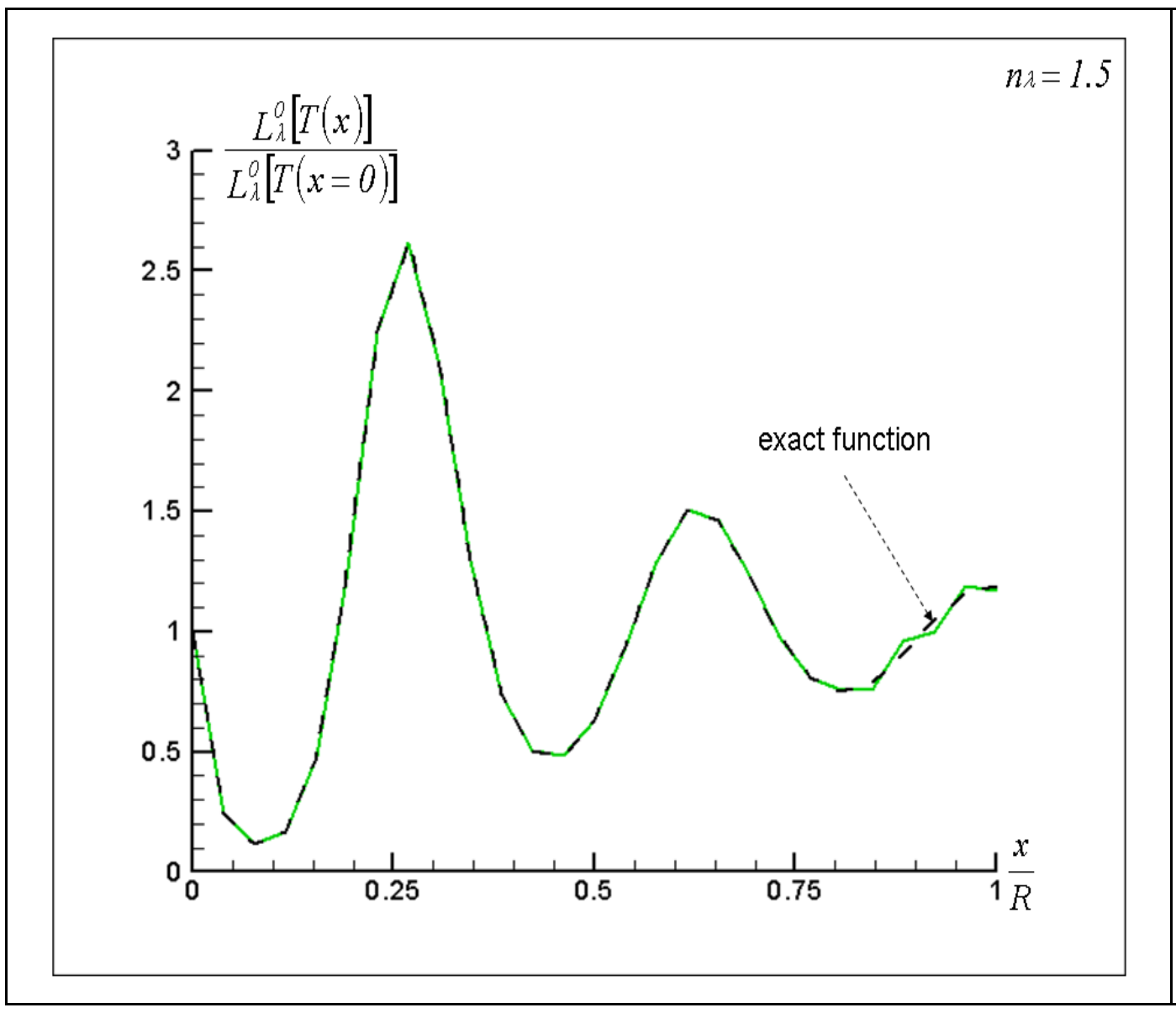

Fig 4a: retrieved Planck function for $N = 27$

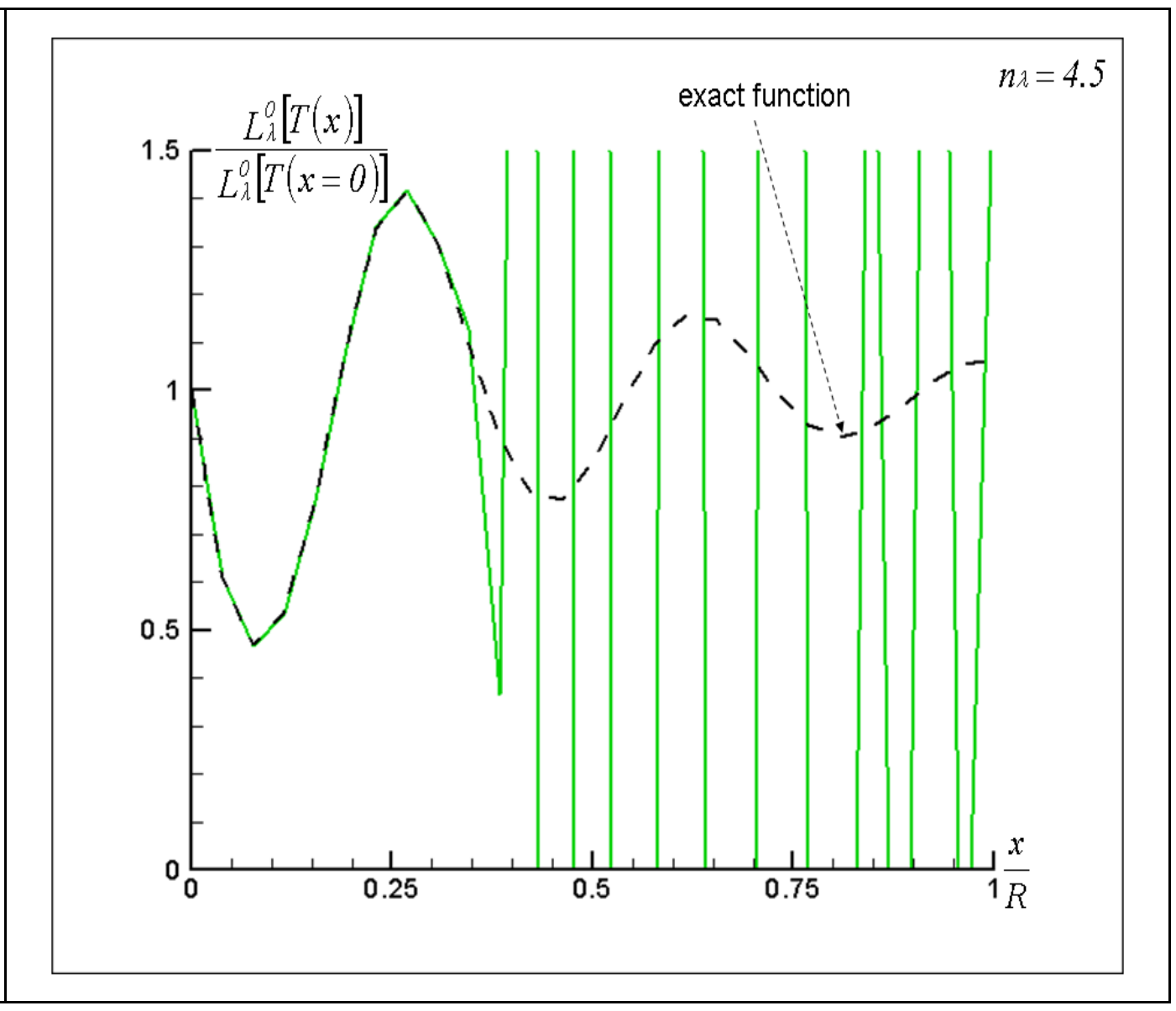

Fig 4b: retrieved Planck function for $N = 27$

Hence it appears that for refraction indices greater than one, an excellent approximation of the temperature field cannot be retrieved everywhere inside the whole cylinder, even for perfect data without any noise, and that for a relatively large refractive index, typically in the range $n_\lambda \in [1.5, 3]$ for many dense media, the temperature field cannot be estimated in an important area of the cylinder when using a simple LU decomposition. This is particularly true for internal temperature fields of complex shape inside highly refracting media, when using a consequent number of data.
The previous numerical inversion procedure has been then applied to a noisy data set with $g(x) = \overline{g}(x)(1 + \delta \langle r \rangle)$, where $\overline{g}(x)$ is a perfect non noisy data set obtained from a direct calculation, $\langle r \rangle$ a random number such that $-1 \le \langle r \rangle \le 1$ and $\delta$ a parameter characterizing the noise intensity. The corresponding results, for the same internal temperature field and thermo-physical constants, are depicted on the two following figures 5a-b for various noise intensity parameters when the number of cells is 100 and 27 corresponding to the above determined threshold value. For a large number of cells, the numerical procedure is unable to produce satisfactory results even for an insignificant value of the noise intensity, while for a number of cells close to the threshold value, the results are acceptable on the partial area $x \leq \frac{R}{n_\lambda}$ for very small additive perturbations. In all cases however the temperature field cannot be here obtained in the particular area $x > \frac{R}{n_\lambda}$.

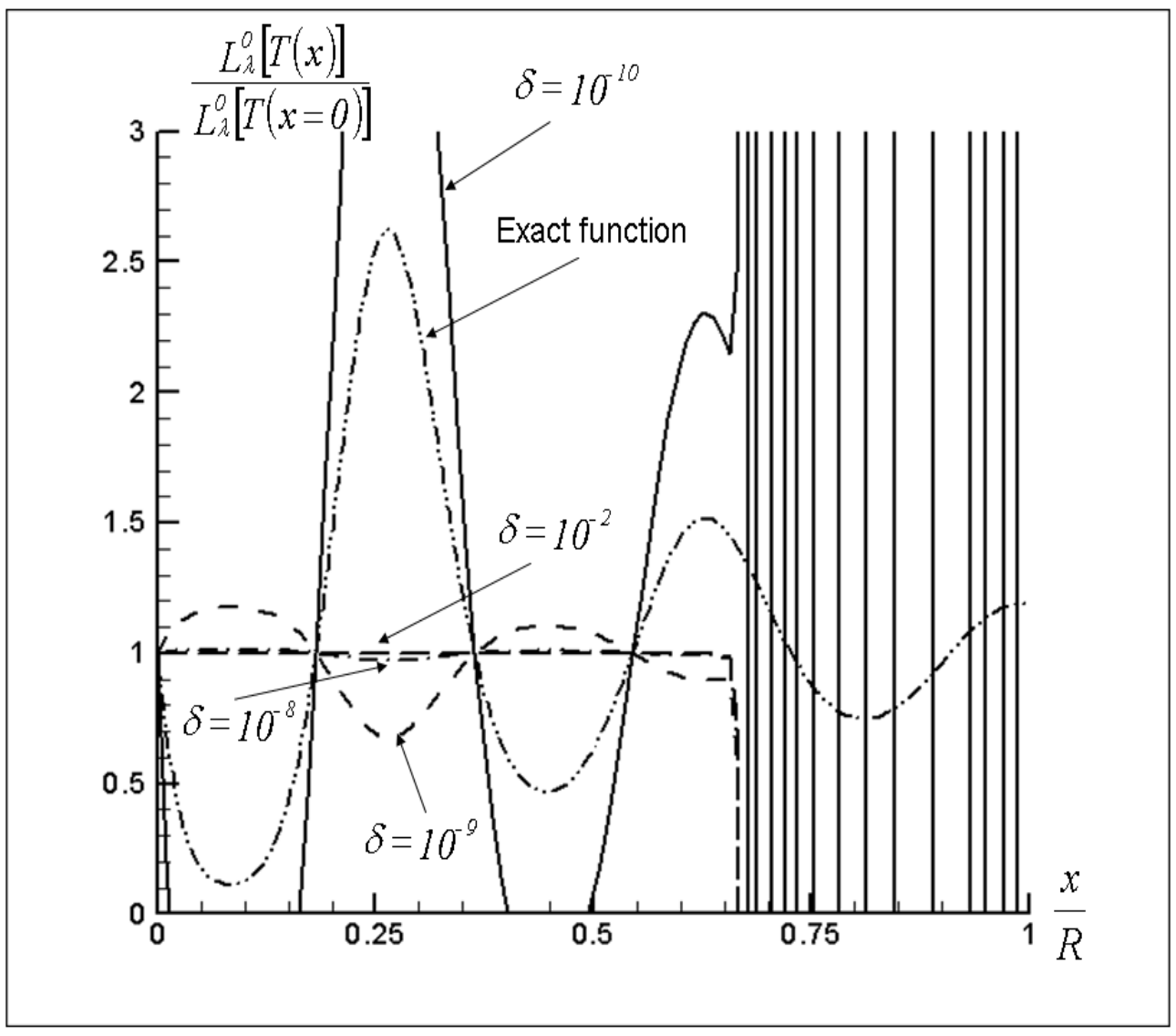

Fig 5a: retrieved Planck function with noisy data for $N = 100$

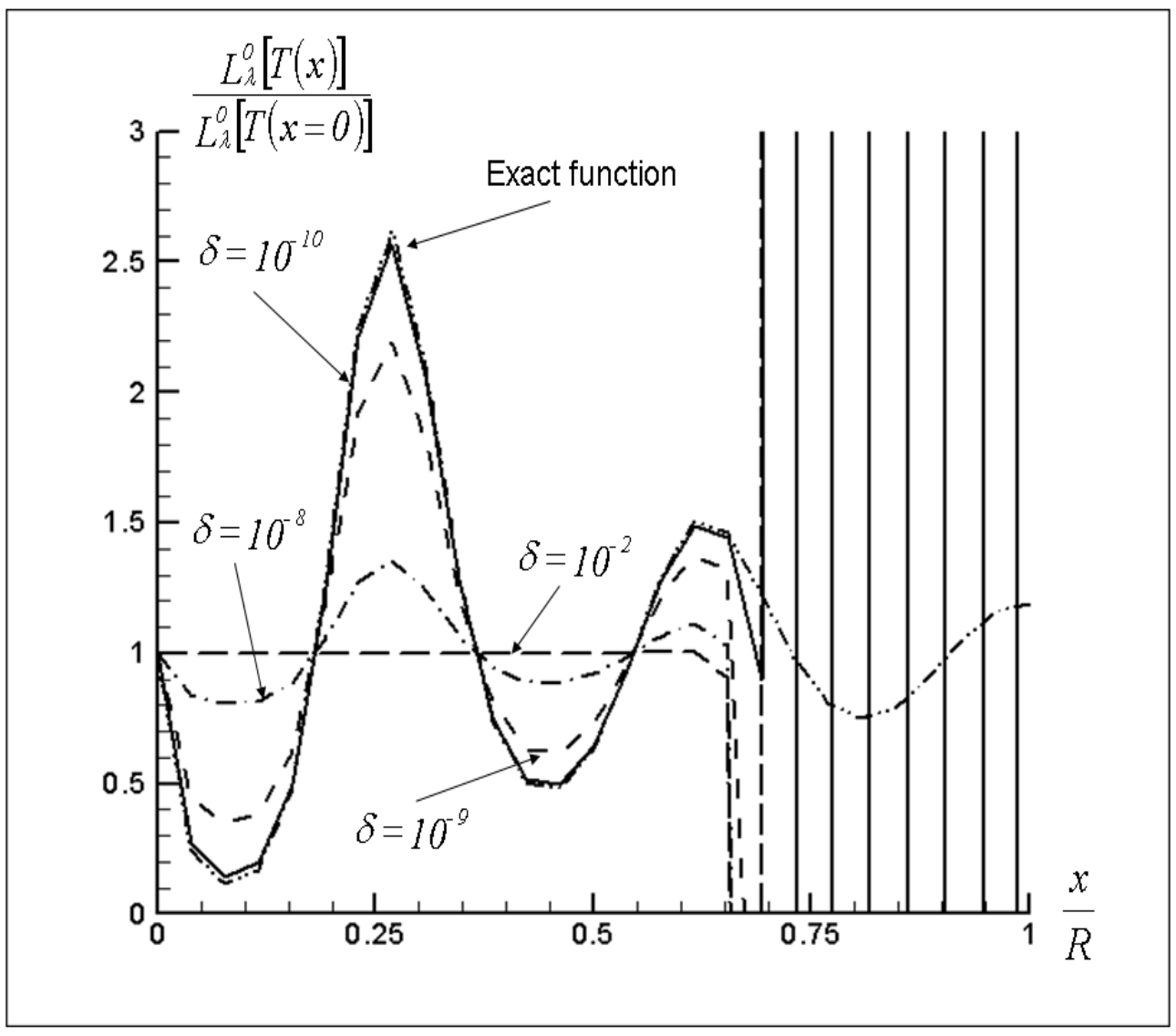

Fig 5b: retrieved Planck function with noisy data for $N = 27$

One has to decide then, before dealing with noisy input *g* values representing a real experimental set of data, if Eq. (2) has a solution for all *x* on the set $[0, R]$ or if, as suggested by the previous numerical study, Eq. (2) can only be inverted in the range $x \leq \frac{R}{n_\lambda}$ and belongs then to the class of missing data problems.

### 3.3 Solution by the SVD decomposition

Applying a singular value decomposition (SVD) of matrix *C*, i.e. $C = U \; diag\,(w_i) V^T$, where *U* and *V* are two orthogonal matrices, the solution of Eq. (5) writes:

$$L_\lambda^0(T) = V \left[ diag\left(\frac{1}{w_i}\right)\right] U^T g \qquad (6)$$

Then, introducing the *C* matrix condition number $cond\; C = \frac{sup\,|w_i|}{min\,|w_i|}$, leads for this example ($n_\lambda = 1.5$) to the results presented on table 1. These results indicate that matrix *C* is highly ill-conditioned for a large number of cells, greater than the step number. Since $C$ is ill-conditioned, a powerful way widely used to obtain an approximate solution is to minimise the residual $\left|C\, L_\lambda^0(T) - g\right|$ by zeroing the small eigenvalues [10, 13, 17]. To do so, one chooses a regularisation parameter $\alpha$ such that if $\left|w_j\right| \leq \alpha \sup \left|w_i\right|$ then $w_j = 0$. For the numerical chosen example, the direct calculation gives a correct solution for $\min \left|w_i\right| \geq 10^{-12}$, so that the optimal $\alpha$ is about $\alpha \approx 10^{-12}$.

| Number of cells | $\min \left\|w_i\right\|$ | $\sup \left\|w_i\right\|$ | cond C |
|---|---|---|---|
| *10* | $9.783\, 10^{-6}$ | *5.598* | $5.722\, 10^{5}$ |
| *20* | $1.132\, 10^{-12}$ | *5.648* | $4.990\, 10^{12}$ |
| *50* | $2.447\, 10^{-17}$ | *5.673* | $2.318\, 10^{16}$ |
| *100* | $9.697\, 10^{-17}$ | *5.681* | $5.858\, 10^{16}$ |
| *500* | $2.420\, 10^{-17}$ | *5.687* | $2.350\, 10^{17}$ |

The approximate solution obtained with a regularisation parameter $\alpha = 10^{-12}$ is depicted on figures 5a and 5b. For refractive indices not too high, exemplified here on Fig. 6a when $n_\lambda = 1.5$, the approximate solution with regularization parameter is very close to the exact one, on the range $r \leq \frac{R}{n_\lambda}$ and also on the set $\left[\frac{R}{n_\lambda}, R\right]$, except at the single point $r = R$. This allows us to admit that, from a practical point of view, a simple discretization of Eq. (2) followed by a singular value decomposition of matrix $C$ and zeroing the small eigenvalues leads to a satisfactory solution by using only directional and no spectral intensities. However, for important refractive indices, and/or important absorption coefficients, as illustrated on Fig. 6b when $n_\lambda = 4.5$, a regularisation procedure gives a good approximation for $r \leq \frac{R}{n_\lambda}$ as when a single LU decomposition is performed, and fails in obtaining a satisfactory solution on $\left[\frac{R}{n_\lambda}, R\right]$ even for large scales of the regularisation parameter. This zeroing procedure works well enough when the number of cells is not too important, which is generally the case in practical cases, but for cylinders of large dimensions with internal temperature fields of complex shapes, for which a significant number of data may be necessary, and if the refractive index is relatively high, the proposed numerical scheme will produce poor quality approximations of the temperature field.

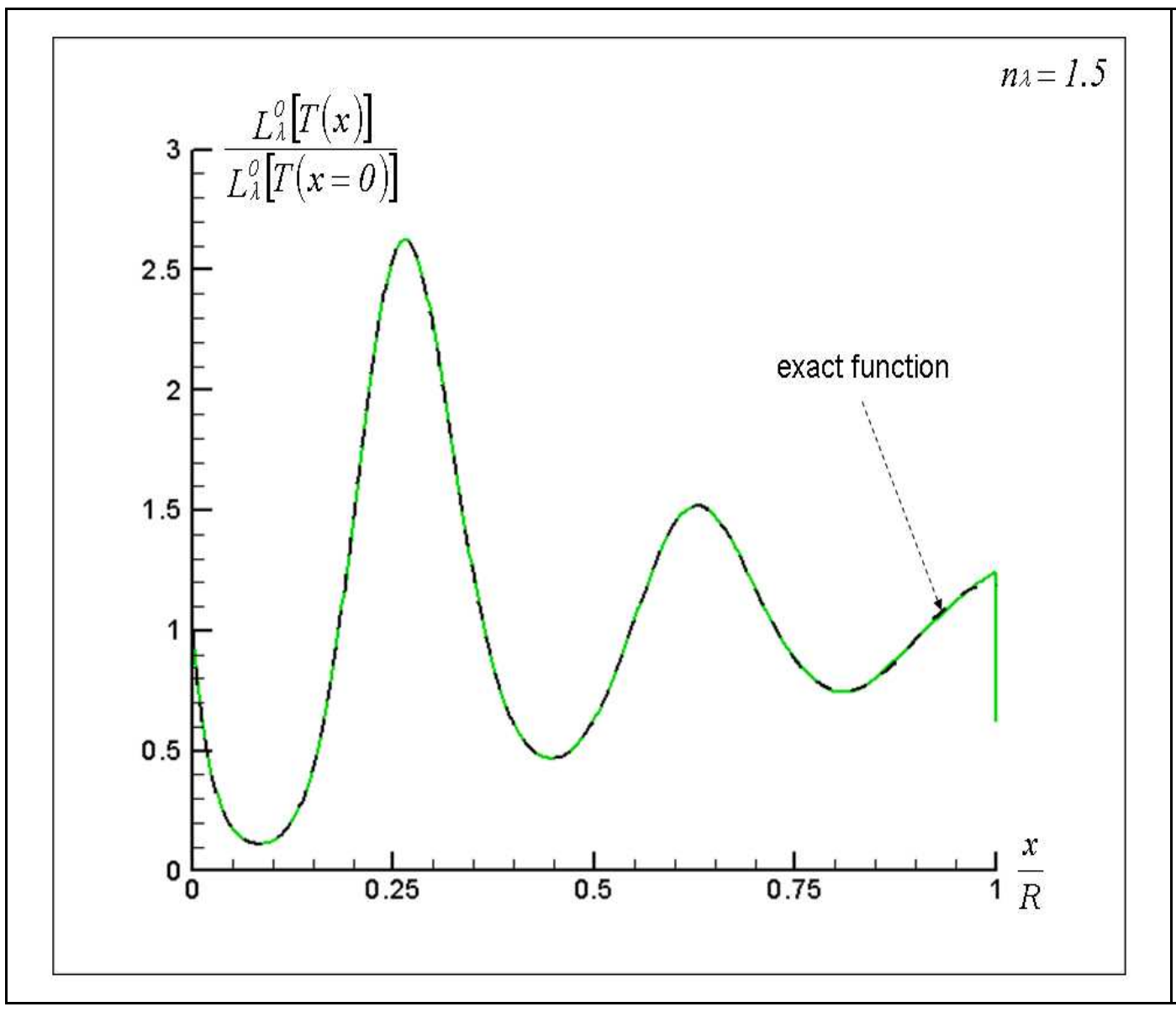

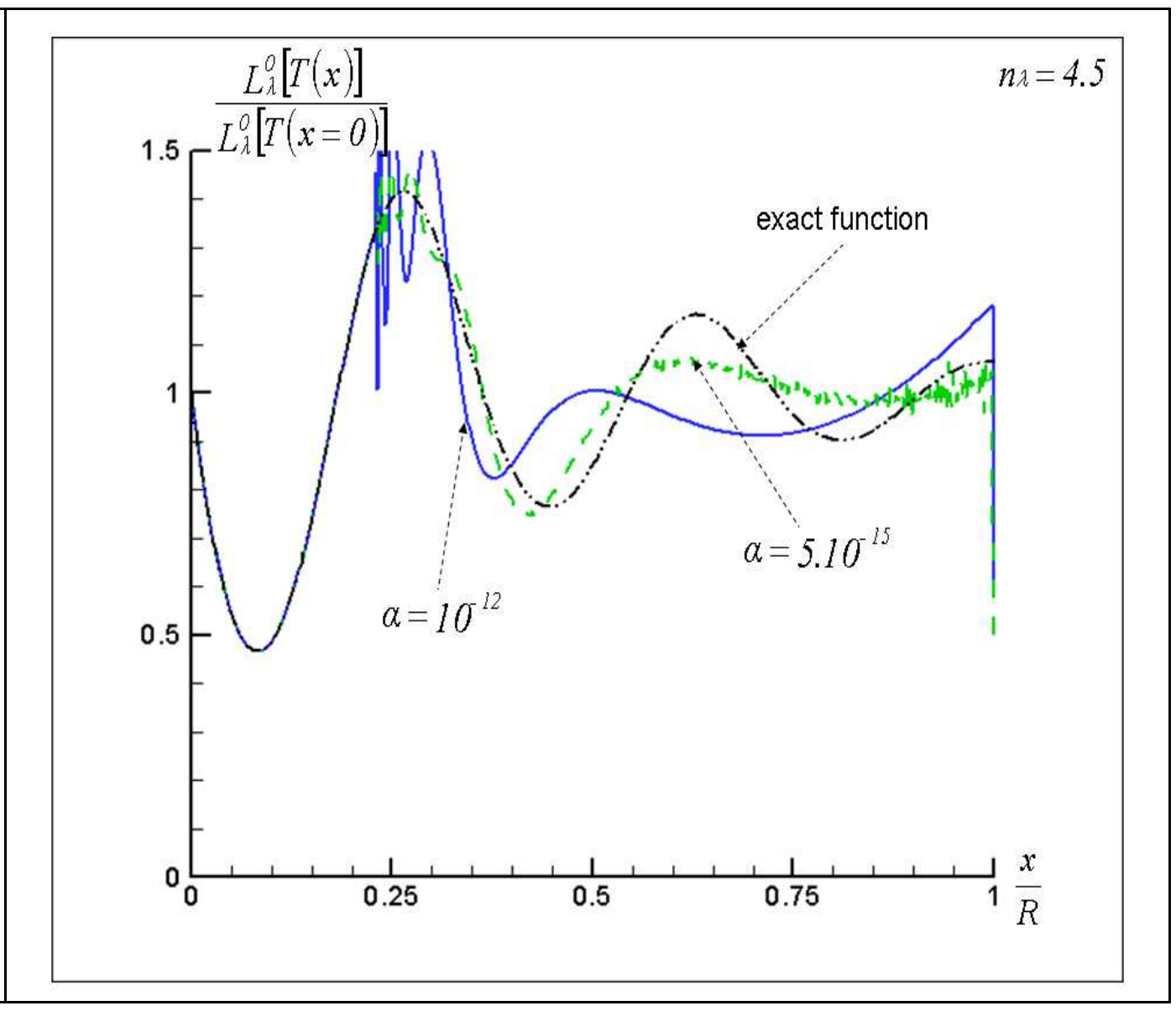

Fig 6a: retrieved Planck function when using a regularization parameter $\alpha = 10^{-12}$, $n_\lambda = 1.5$

Fig 6b: retrieved Planck function when using a regularization parameter, $n_\lambda = 4.5$

A SVD has also been performed for a cylinder of 24 cm radius filled with a STM of absorption coefficient $\kappa_\lambda = 10\ m^{-1}$ and refractive index $n_\lambda = 1.5$, submitted to a sinusoidal temperature field, with noisy data for two different cells numbers when the noise intensity is 1% (i.e. $\delta = 10^{-2}$). As illustrated on the following figures 7a-b, a SVD gives poor quality results even for relatively high regularization coefficients, especially for a large number of cells. When this number tends towards the threshold value, the results are more accurate and acceptable, also for important regularization factors, in the range $r \leq \frac{R}{n_\lambda}$, but remain of poor quality when $x > \frac{R}{n_\lambda}$. This latter remark enhances that the restitution of the temperature field inside a cylinder of dense STM with high refractive indices suffers from two major restrictions: 1) due to the non unit refractive index, the restitution cannot be performed on the range $x > \frac{R}{n_\lambda}$ without any regularization procedure on the governing operator, even for perfect non noisy data, and 2) the presence of weakly noisy data significantly alters the above operator regularisation procedure in such a way that it necessarily implies an initial preconditioning of the useful perturbed data.

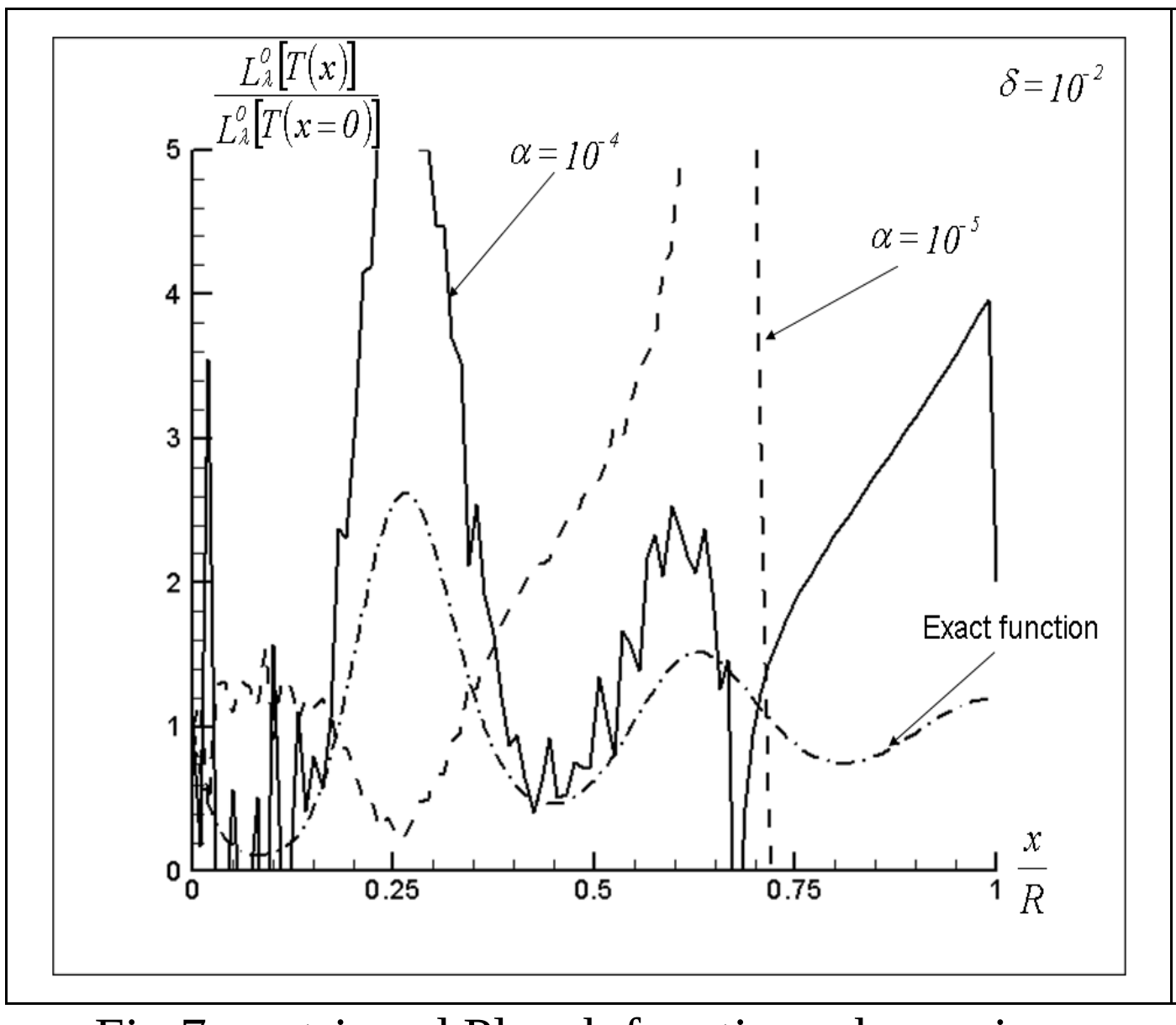

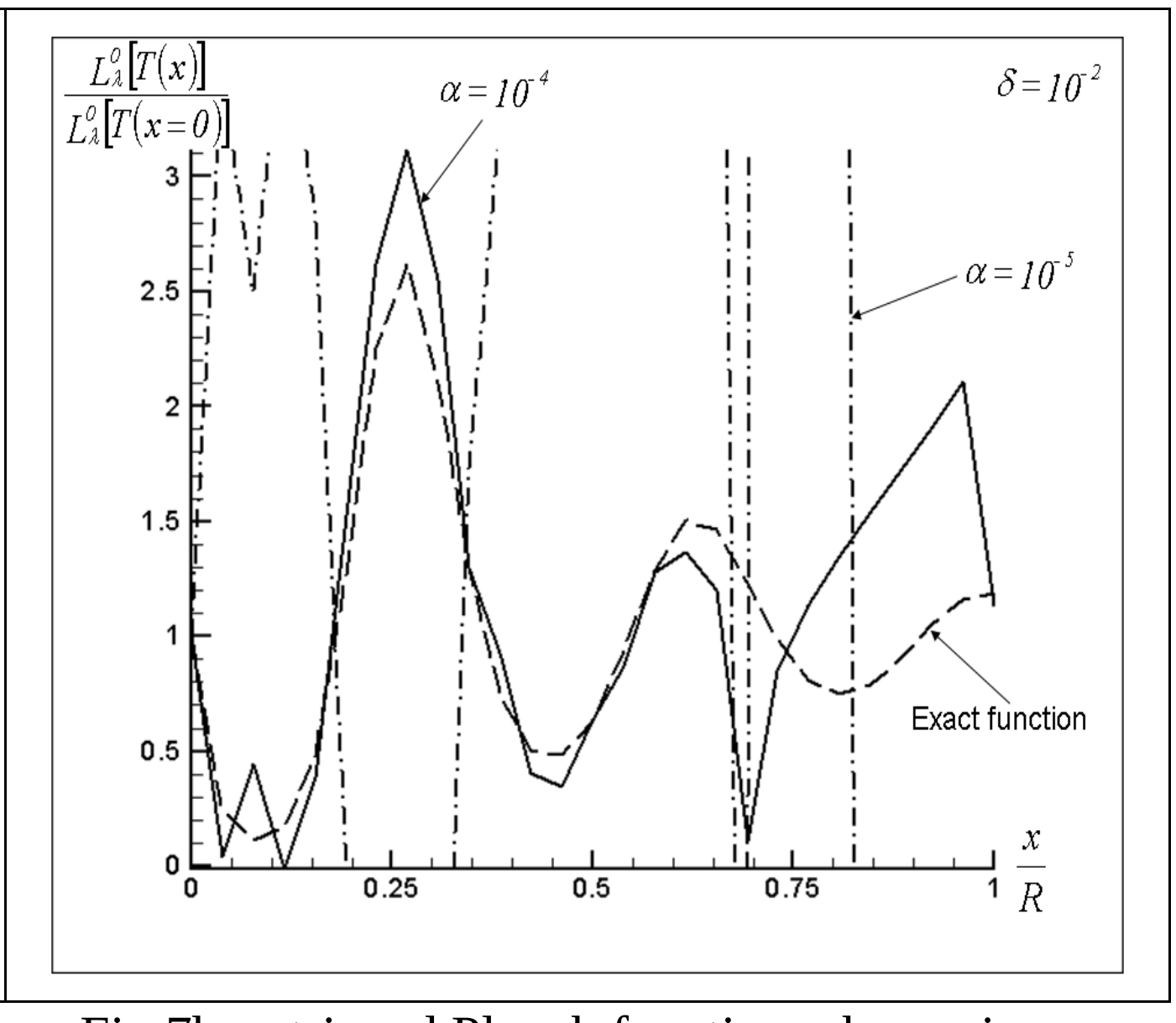

Fig 7a: retrieved Planck function when using a

Fig 7b: retrieved Planck function when using a regularization parameter, $N = 100$ regularization parameter, $N = 27$

To avoid at this stage a discussion on the possible techniques of data conditioning such as filtering, and moreover understand why such an operator regularisation even for non noisy data, is necessary, i.e. why a direct LU decomposition gives only a good solution on the range $r \le \frac{R}{n_\lambda}$, we shall determine in the next subsection the exact solution of the integral equation (2) when no perturbation is added to a perfect set of emerging intensity data. If, as suggested by a simple SVD for perfect non noisy data, there is an exact theoretical solution on the whole set $0 \le r \le R$, this opens the way to studies whose scope is the complete restitution of a temperature field inside a cylinder filed with a dense STM in presence of noisy directional emerging intensity data from an experimental apparatus.

## 4. EXACT SOLUTION OF THE INTEGRAL EQUATION

### 4.1 Formal exact solution

By using the following variables change

$$\begin{aligned} u &= \tau^2 \quad ; \quad y = \frac{\kappa_\lambda^2 x^2}{n_\lambda^2} \\ v &= \tau_0^2 - u \quad ; \quad z = \tau_0^2 - y \end{aligned} \tag{7}$$

and introducing functions *F* and *G* defined by

$$F(v) = L_\lambda^0\left[T\left(\frac{\sqrt{\tau_0^2 - v}}{\kappa_\lambda}\right)\right] \quad ; \quad G(z) = 2\,g\left(\frac{n_\lambda\sqrt{\tau_0^2 - z}}{\kappa_\lambda}\right) \tag{8}$$

one obtains

$$\int_{v=0}^{z} F(v)\frac{\cosh\left(\sqrt{z-v}\right)}{\sqrt{z-v}}\,dv = G(z) \tag{9}$$

In Eq. (9), $z \in [z_1, z_2] = \left[\left(1 - \frac{1}{n_\lambda^2}\right)\tau_0^2, \tau_0^2\right]$ can never reach zero if the refraction index is different from 1, and the classical theorems on the Laplace transform can no longer be applied. To avoid this difficulty, an origin translation is applied by introducing $z^* = z - z_1 \in [z^*_1, z^*_2] = \left[0, \frac{\tau_0^2}{n_\lambda^2}\right]$, from which one deduces

$$\int_{u^*=0}^{z^*} J(u^*)\frac{\cosh\left(\sqrt{z^* - u^*}\right)}{\sqrt{z^* - u^*}}\,du^* = \tilde{H}(z^*) \tag{10}$$

with $u^* = v - z_1$, $F(u^* + z_1) = J(u^*)$ and $\tilde{H}(z^*) = G(z) - \int_{u^*=0}^{z_1} J(-u^*)\frac{\cosh\left(\sqrt{z^* + u^*}\right)}{\sqrt{z^* + u^*}}\,du^*$.

The boundary values of these functions can be written explicitly:

$$J(0)=L_\lambda^0\left[T\left(\frac{R}{n_\lambda}\right)\right] \quad ; \qquad J(z^*_2)=L_\lambda^0[T(0)]$$

$$\tilde{H}(0)=0 \qquad ; \quad \tilde{H}(z^*_2)=2\,\kappa_\lambda \int_{r=0}^{\frac{R}{n_\lambda}} L_\lambda^0[T(r)]\cosh(\kappa_\lambda r)\,dr \tag{11}$$

Eq. (10) is a Faltung convolution equation of kernel $K(t^*)=\frac{\cosh\left(\sqrt{t^*}\right)}{\sqrt{t^*}}$, whose Laplace transform is $[L(K)](p^*)=\sqrt{\frac{\pi}{p^*}}\,e^{\frac{1}{4p^*}}$. Applying a Laplace transform to Eq. (10) leads to:

$$L_\lambda^0\left[T\left(\frac{x}{n_\lambda}\right)\right]=\frac{1}{\pi}\int_{v=0}^{z^*}\tilde{H}'(v)\frac{\cos\left(\sqrt{z^*-v}\right)}{\sqrt{z^*-v}}\,dv \tag{12}$$

With the help of the previous variables and functions changes, the right member of Eq. (10) can be reformulated under the exact form:

$$\int_{v=0}^{z^*}\tilde{H}'(v)\frac{\cos\left(\sqrt{z^*-v}\right)}{\sqrt{z^*-v}}\,dv=-\frac{1}{\kappa_\lambda}\int_{r=\frac{x^*}{n_\lambda}}^{R}\frac{d}{dr}\left\{\tilde{H}\left[\kappa_\lambda^2\left(R^2-r^2\right)\right]\right\}\frac{\cos\left(\kappa_\lambda\sqrt{r^2-\frac{x^{*2}}{n_\lambda^2}}\right)}{\sqrt{r^2-\frac{x^{*2}}{n_\lambda^2}}}\,dr \tag{13}$$

where $\frac{x^*}{n_\lambda}=\sqrt{\frac{x^2}{n_\lambda^2}+\left(1-\frac{1}{n_\lambda^2}\right)R^2}$ . Replacing $\tilde{H}$ by its value in terms of $g$ and $L_\lambda^0$ functions and using the two successive variables changes defined by $r^*=\sqrt{r^2-\left(1-\frac{1}{n_\lambda^2}\right)R^2}$ and $s=n_\lambda r^*$, one finally obtains (after a straightforward calculation) :

$$L_\lambda^0\left[T\left(\frac{x}{n_\lambda}\right)\right]-\frac{2\,n_\lambda^2}{\pi}\int_{y=\frac{R}{n_\lambda}}^{R}\frac{y\,L_\lambda^0[T(y)]}{y^2-\frac{x^2}{n_\lambda^2}}\left[\begin{array}{l}\frac{\sqrt{R^2-x^2}}{n_\lambda\sqrt{y^2-\frac{R^2}{n_\lambda^2}}}\cos\left(\frac{\kappa_\lambda}{n_\lambda}\sqrt{R^2-x^2}\right)\cosh\left(\kappa_\lambda\sqrt{y^2-\frac{R^2}{n_\lambda^2}}\right)\\ -\sin\left(\frac{\kappa_\lambda}{n_\lambda}\sqrt{R^2-x^2}\right)\sinh\left(\kappa_\lambda\sqrt{y^2-\frac{R^2}{n_\lambda^2}}\right)\end{array}\right]dy$$

$$=-\frac{2\,n_\lambda}{\pi\,\kappa_\lambda}\int_{s=x}^{R}\frac{g'(s)\cos\left(\frac{\kappa_\lambda}{n_\lambda}\sqrt{s^2-x^2}\right)}{\sqrt{s^2-x^2}}\,ds \tag{14}$$

where $g'$ is the derivative of the data set $g$. Equation (14) is the exact inverse solution of equation (2).

Note that the complete data set $g$ on the whole set $[0,R]$ must be used to compute the temperature field in the zone $\left[0,\frac{R}{n_\lambda}\right]$, and that this field depends on the temperature field in the region $\left[\frac{R}{n_\lambda},R\right]$. Under this form, Eq. (14) looks like a 2nd kind Fredholm equation, but however is not such an integral equation. Indeed, the previous equation is strictly equivalent to the analogous form

$$\Phi(r) - \lambda \int_{y=\frac{R}{n_\lambda}}^{R} \Phi(y)\,K(r,y)\,dy = F(r) \tag{15}$$

where $\Phi$ is the unknown function defined on $r \in \left[0,\frac{R}{n_\lambda}\right]$ for the first LHS member, while the definition interval of the integral is $\left[\frac{R}{n_\lambda},R\right]$. One can notice that from the exact definition of the $g$ function given by Eq. (2), Eq. (14) can be replaced by the strictly equivalent equation

$$L_\lambda^0\left[T\left(\frac{x}{n_\lambda}\right)\right] = -\frac{2\,n_\lambda}{\pi\,\kappa_\lambda}\int_{s=x}^{R} \frac{h'(s)\cos\left(\frac{\kappa_\lambda}{n_\lambda}\sqrt{s^2-x^2}\right)}{\sqrt{s^2-x^2}}\,ds \tag{16}$$

where the $h$ function is defined by

$$h(x) = \int_{r=x}^{R} \frac{r\,\kappa_\lambda\,L_\lambda^0\left[T\left(\frac{r}{n_\lambda}\right)\right]}{n_\lambda\sqrt{r^2-x^2}}\cosh\left(\frac{\kappa_\lambda}{n_\lambda}\sqrt{r^2-x^2}\right)dr \tag{17}$$

Function $h$ is the generalised $g$ function for $n_\lambda \neq 1$ and equivalent to $g$ when $n_\lambda = 1$.

Under this latter form, it is therefore obvious that the inversion of Eq. (2) only leads to a partial temperature field on $\left[0,\frac{R}{n_\lambda}\right]$. Note also that function $h$ cannot be simply related to the experimental set $g$ and cannot be applied at this stage from a practical point of view to determine the temperature field.

### 4.2 Numerical verification

From the exact expression (17) of function $h$ it is easy to prove that Eq. (14) is verified at $x = R$. We present in this section the results obtained by solving numerically equation (14) with the numerical scheme presented in annex 1.

The restitution of the temperature field, so as the evolution of the function $h$, analogous to the data $g$ set for unit refractive indices, are presented on figures 8a-d for $n_\lambda = 1.5$ and $n_\lambda = 4.5$, for exact $h$ data and for "data" obtained from a calculated $g$ set by a SVD of matrix $C$ followed by a simple zeroing with $\alpha = 10^{-12}$, when no perturbation is done on the exact $g$ data. The numerical computation of Eq. (14) gives excellent results compared to the exact ones, even if the temperature field is not correctly estimated in the neighbourhood of $x = \frac{R}{n_\lambda}$, due to the fact that $h'_N$ is forced to have a finite value.

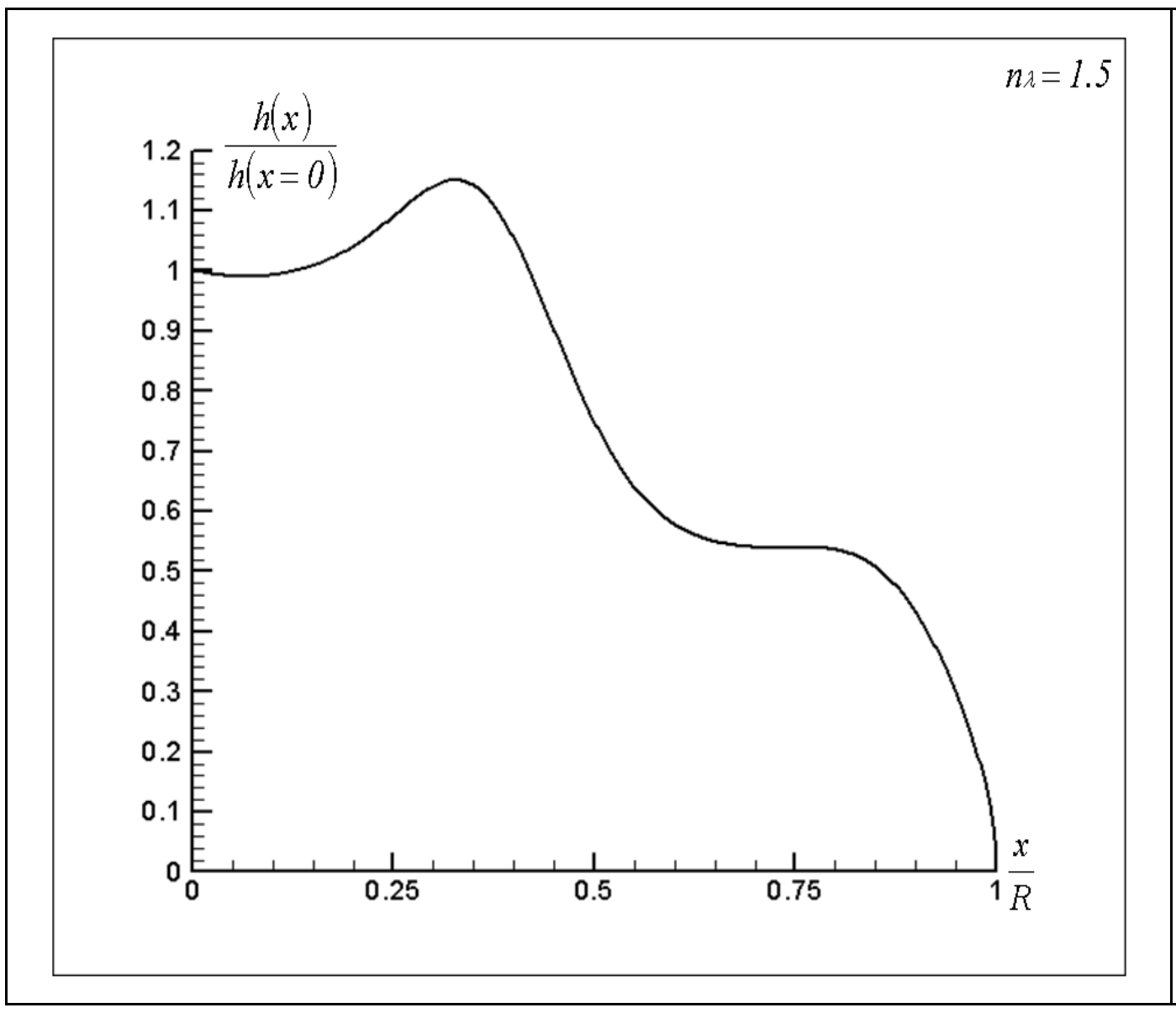

Fig 8a: evolution of function $h$ for $x \le R$, $n_\lambda = 1.5$

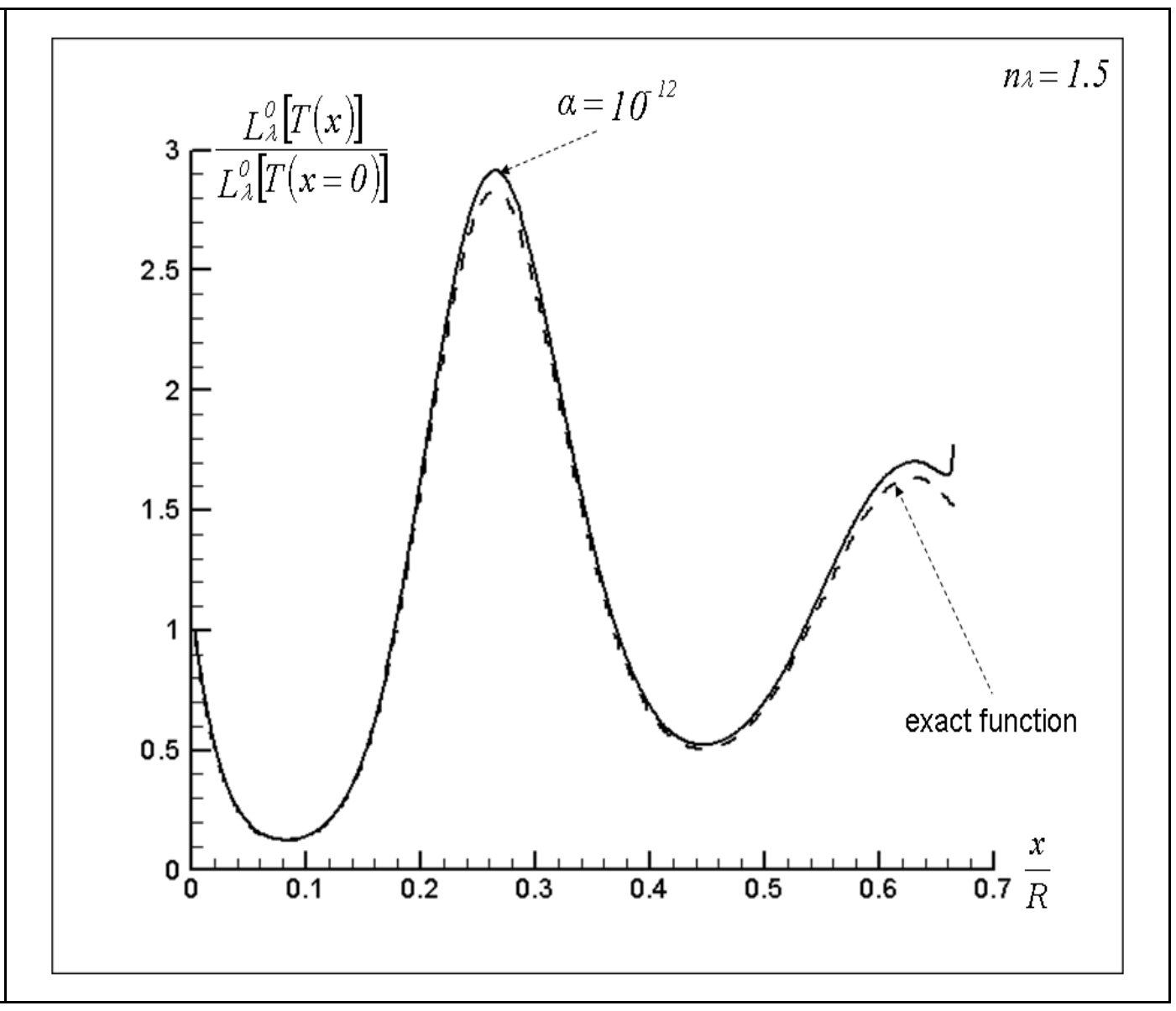

Fig 8b: retrieved Planck function with Eq. (14) on the range $\left[0, \frac{R}{n_\lambda}\right]$ for $n_\lambda = 1.5$

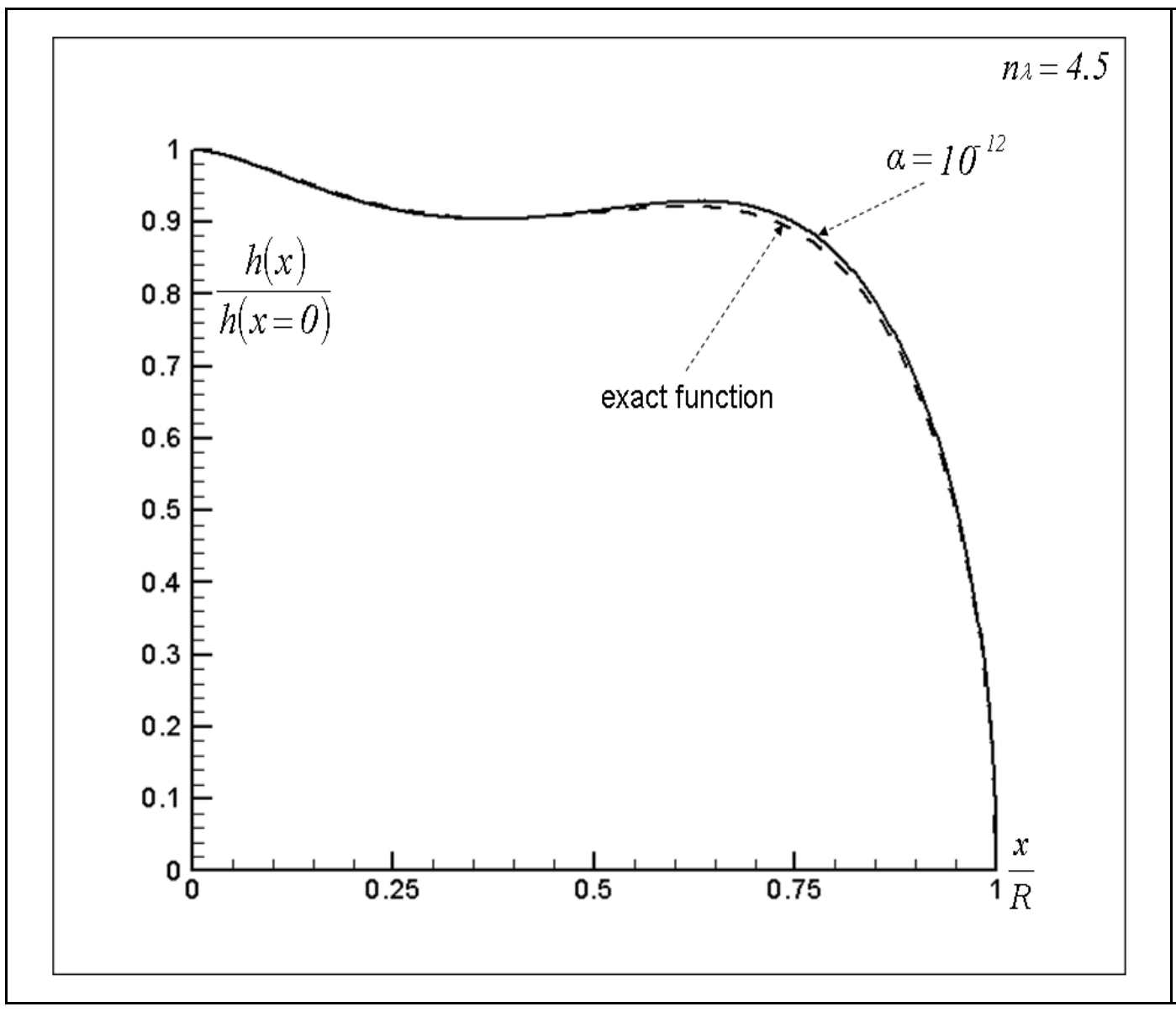

Fig 8c: evolution of function $h$ for $x \le R$, $n_\lambda = 4.5$

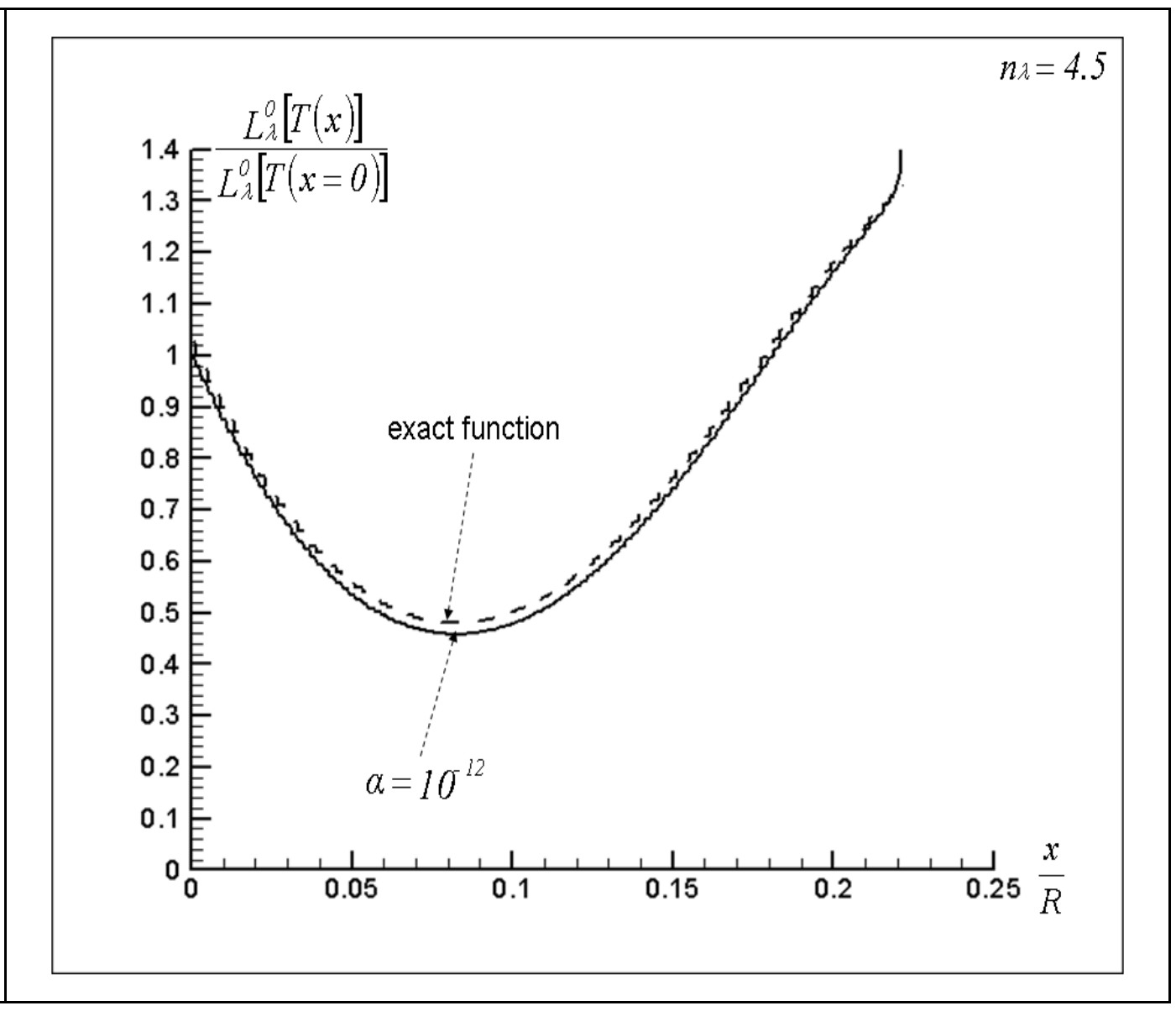

Fig 8d: retrieved Planck function with Eq. (14) on the range $\left[0, \frac{R}{n_\lambda}\right]$ for $n_\lambda = 4.5$

The main interest of this operation is not here to retrieve the temperature field on $\left[0,\frac{R}{n_\lambda}\right]$, but to calculate the generalised $h$ function on $[0,R]$, so that the partial temperature field on $\left[\frac{R}{n_\lambda},R\right]$ is governed by the integral equation

$$\int_{r=\frac{R}{n_\lambda}}^{R} \frac{r\,\kappa_\lambda\,L_\lambda^0[T(r)]}{\sqrt{r^2-\frac{x^2}{n_\lambda^2}}} \cosh\left(\kappa_\lambda\sqrt{r^2-\frac{x^2}{n_\lambda^2}}\right) dr = g(x)-h(x)=\psi(x), \qquad (18)$$

Function $\psi$ is completely known on $[0,R]$ so that Eq. (18) can be considered as a first kind Fredholm equation for $x\in\left[\frac{R}{n_\lambda},R\right]$, where $\psi$ is the data function of the equation, $K(r,x)=\frac{r\cosh\left(\kappa_\lambda\sqrt{r^2-\frac{x^2}{n_\lambda^2}}\right)}{\sqrt{r^2-\frac{x^2}{n_\lambda^2}}}$ is the kernel of the integral equation and $L_\lambda^0[T(r)]$ is the unknown function to retrieve on $\left[\frac{R}{n_\lambda},R\right]$. Equation (18) is only depending on the unknown part of the temperature field, which means that if this latter equation has a solution, then the global problem summed up by the general equation (2) does not belong to the class of "missing data problems", as frequently mentioned, and that there is hope to find approximate numerical procedures allowing the restitution of the complete temperature field on the whole set $[0,R]$. If however Eq. (18) has no physical admissible solution, directional intensities data are unable to give a complete description of the temperature field inside the cylinder.

## 4.3 Discussion on the Kernels of the operator

The detailed behaviour of the exact $\psi$ function with respect to the absorption coefficient is depicted on the figures 9a-b .

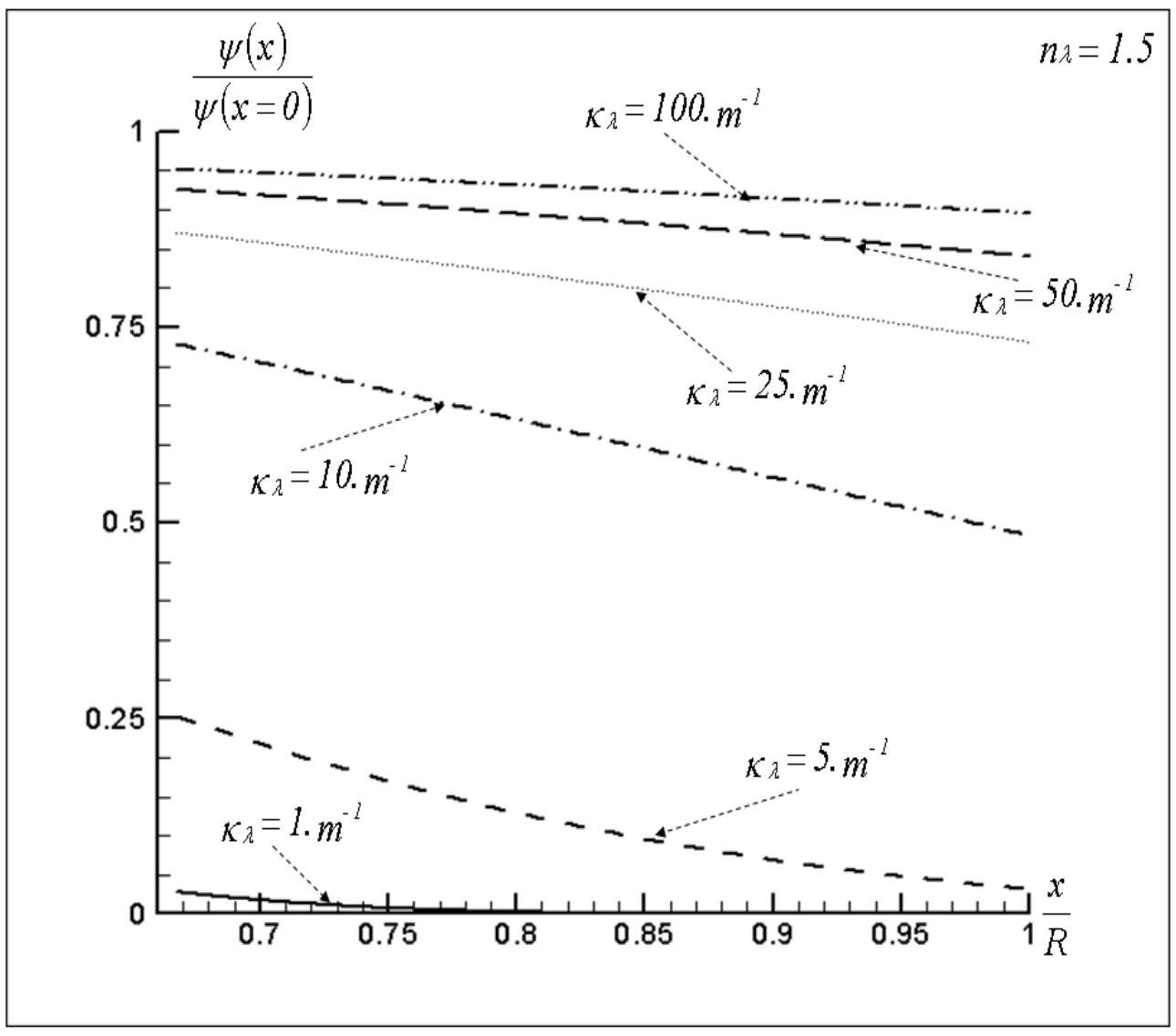

Fig 9a: evolution of function $\psi$ on $\left[\frac{R}{n_\lambda}, R\right]$ for various absorption coefficients, $n_\lambda = 1.5$

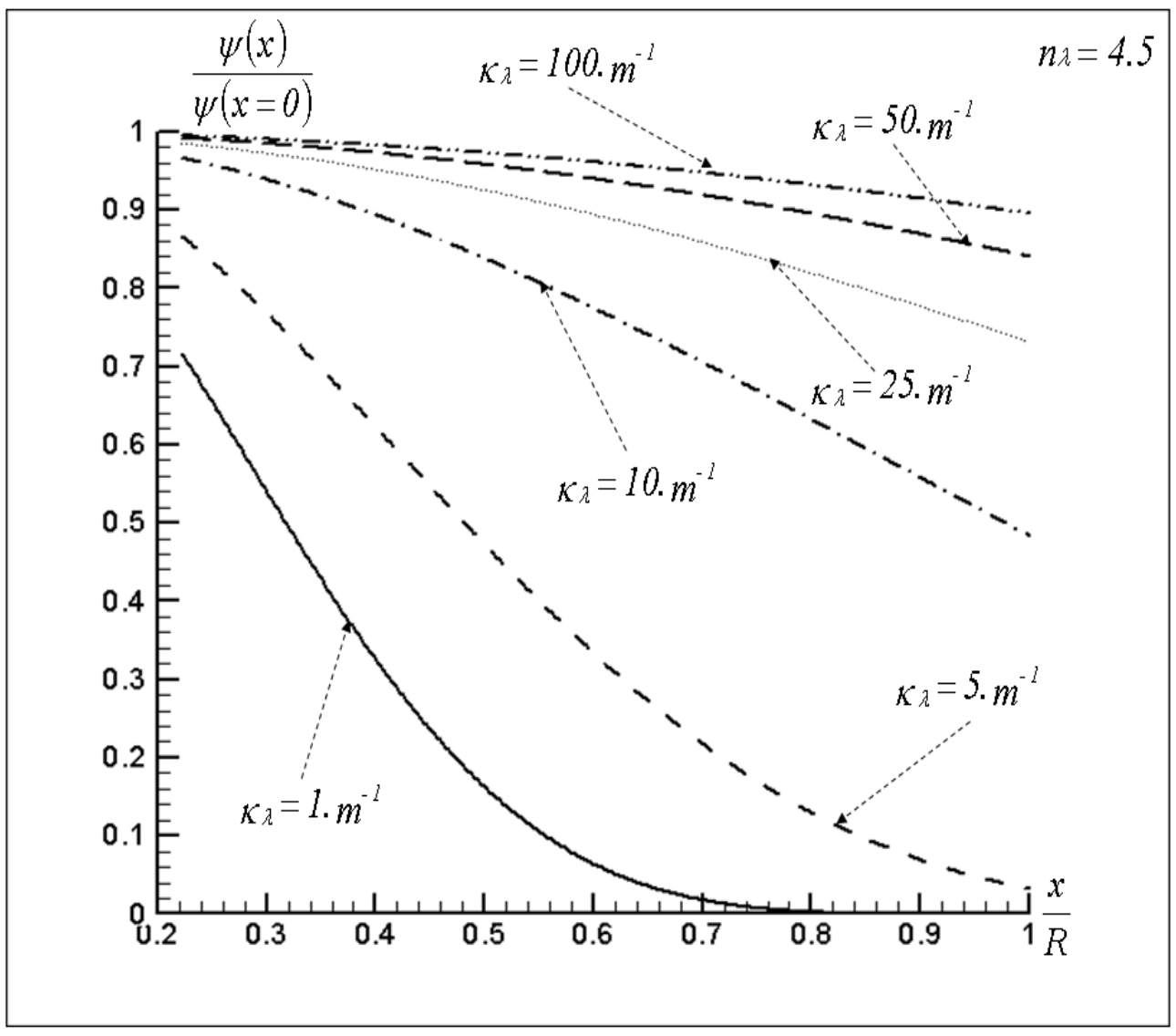

Fig 9b: evolution of function $\psi$ on $\left[\frac{R}{n_\lambda}, R\right]$ for various absorption coefficients, $n_\lambda = 4.5$

For moderate refractive indices, function $\psi$ is approximately linear, of quasi-constant value as soon as the absorption coefficient is greater than $30\,m^{-1}$, and of very low amplitude for weak absorption coefficients. Nevertheless, the sensibility of $\psi$ on the useful set $\left[\frac{R}{n_\lambda}, R\right]$ is poor with respect to the internal position *x*. For higher refractive indices, function $\psi$ remains almost constant for high absorption coefficients, with apparently better sensibility with respect to the internal position.

Before examining the particular properties of Eq. (18), it is worth noting that its discrete form can be written as:

* if $t \leq N-1$ :

$$L^0_\lambda(T_t)\left[sh\left(\kappa_\lambda\sqrt{r^2-\frac{x_j^2}{n_\lambda^2}}\right)\right]_{\frac{R}{n_\lambda}}^{x_t+\frac{\Delta r}{2}} + \sum_{k=t+1}^{N-1} L^0_\lambda(T_k)\left[sh\left(\kappa_\lambda\sqrt{r^2-\frac{x_j^2}{n_\lambda^2}}\right)\right]_{x_k-\frac{\Delta r}{2}}^{x_k+\frac{\Delta r}{2}} \qquad 1 \leq j \leq N$$

$$+ L^0_\lambda(T_N)\left[sh\left(\kappa_\lambda\sqrt{r^2-\frac{x_j^2}{n_\lambda^2}}\right)\right]_{R-\frac{\Delta r}{2}}^{R} = \psi(x_j)$$

(19)

* if $t = N$ :

$$L^0_\lambda(T_N)\left[sh\left(\kappa_\lambda\sqrt{r^2-\frac{x_j^2}{n_\lambda^2}}\right)\right]_{\frac{R}{n_\lambda}}^{R} = \psi(x_j) \qquad 1 \leq j \leq N$$

For $j = N$ the solution is: $$L_\lambda^0(T_N) = \frac{g(R)}{sh\left(\kappa_\lambda R\sqrt{1-\frac{1}{n_\lambda^2}}\right)}$$

When $n_\lambda$ is very close to 1, or equivalently when the number of cells is small, Eq. (19) gives the exact value $T_N$. Since one must seek the exact solution on $\left[\frac{R}{n_\lambda}, R\right]$, the discrete numerical approach involves the grid labelled $t \le j \le N$ and leads to the resolution of a linear system of $N-t+1$ equations with $N-t+1$ unknown quantities, formally written as $\tilde{C}\, L_\lambda^0(T) = \psi$. We present on Figures 10-a and 10-b the numerical results of such a calculation, when a SVD followed by a zeroing is performed on the matrix $\tilde{C}$ and when the number of cells is exactly $N-t+1$, with $N = 100$. For a refractive index $n_\lambda = 1.5$, this leads to $N-t+1=34$, and for $n_\lambda = 4.5$, one has $N-t+1=78$. As it can be seen, there is a slight improvement (relatively to a global discretization of Eq. 2) when Eq. (18) is discretized and a SVD is applied. A numerical solution, although of poor quality, can be exhibited, which let us believe that the integral equation on the partial set $\left[\frac{R}{n_\lambda}, R\right]$ has an exact solution.

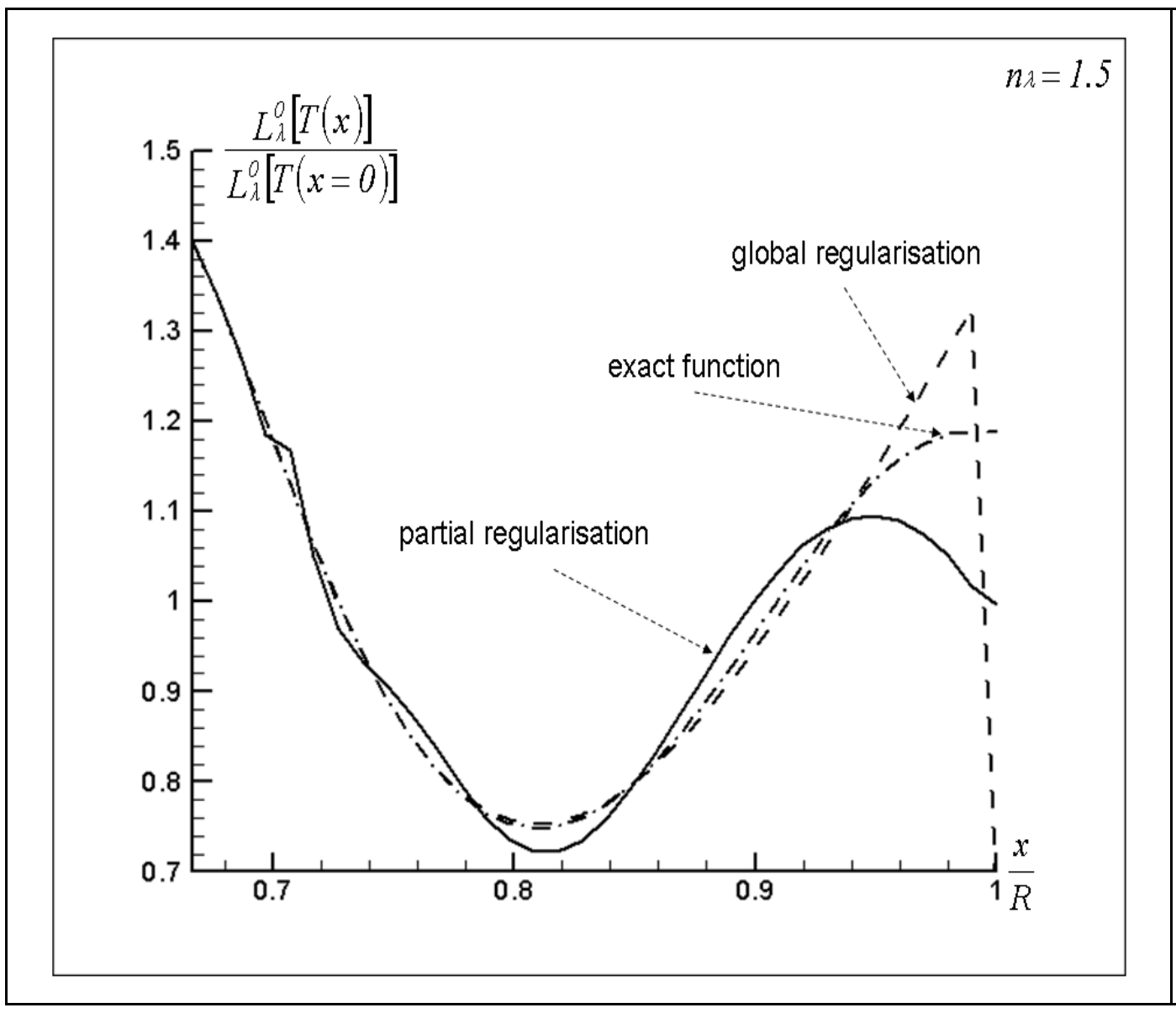

Fig 10a: retrieved Planck function when using a regularization parameter $\alpha = 10^{-9}$, $n_\lambda = 1.5$

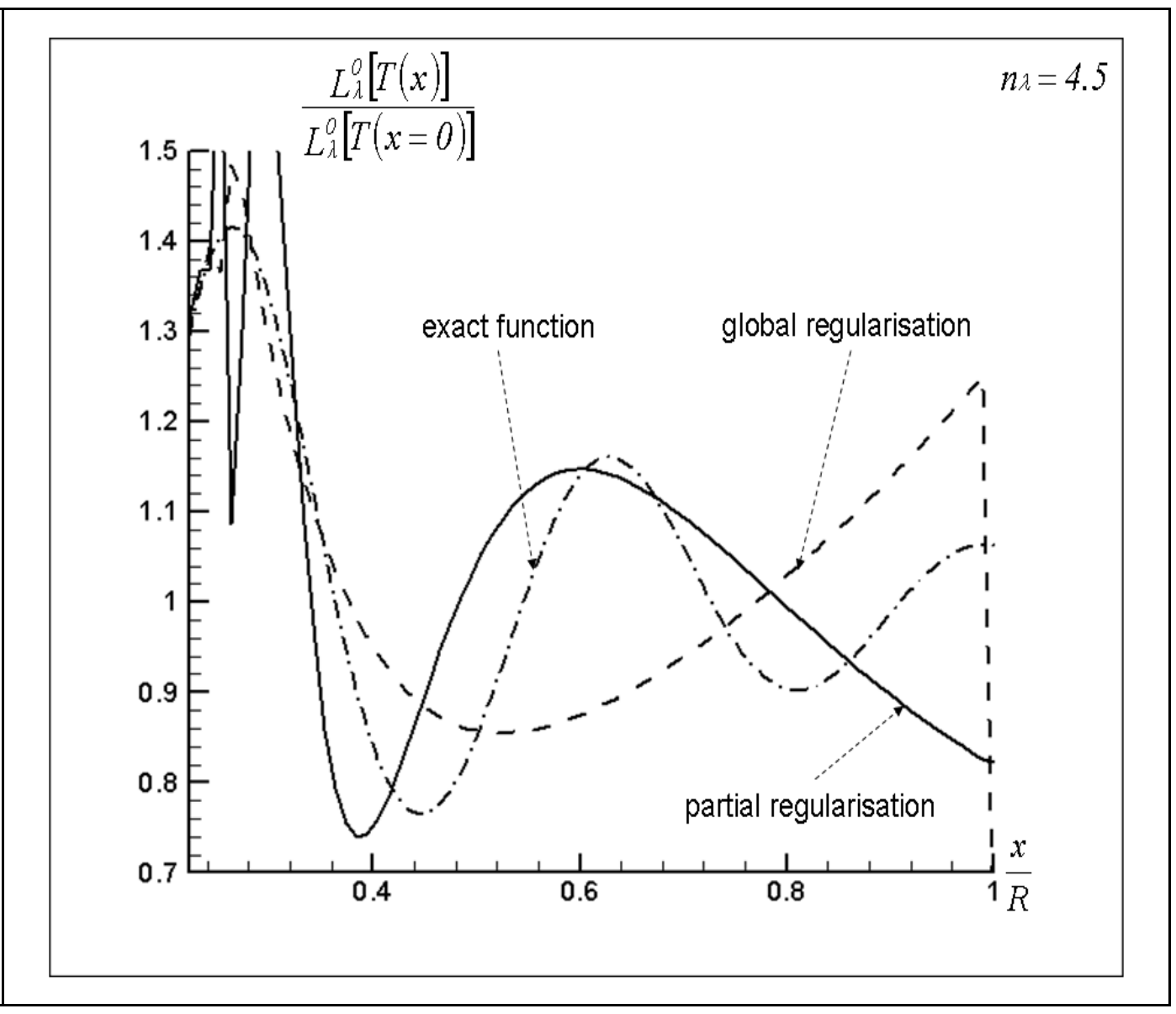

Fig 10b: retrieved Planck function when using a regularization parameter, $\alpha = 10^{-9}$ $n_\lambda = 4.5$

With the notations used for the Laplace transform, Eq. (18) can be rewritten as

$$\int_{v=0}^{\left(1-\frac{1}{n_\lambda^2}\right)\tau_0^2} L_\lambda^0\left[T\left(\frac{\sqrt{\tau_0^2 - v}}{\kappa_\lambda}\right)\right] \frac{ch\left(\sqrt{z-v}\right)}{\sqrt{z-v}}\, dv = 2\,\psi\left(\frac{n_\lambda}{\kappa_\lambda}\sqrt{\tau_0^2 - z}\right) \tag{20}$$

The variable $z$ is such that $z \in \left[\left(1-\frac{1}{n_\lambda^2}\right)\tau_0^2, \left(1-\frac{1}{n_\lambda^4}\right)\tau_0^2\right] \not\subset \left[0, \left(1-\frac{1}{n_\lambda^2}\right)\tau_0^2\right]$. Introducing the new parameter and variables $\alpha^* = \tau_0\sqrt{1-\frac{1}{n_\lambda^2}}$, $w = 1-\frac{v}{\alpha^{*2}} \in [0,1]$ and $z' = n_\lambda^2\left(\frac{z}{\alpha^{*2}}-1\right) \in [0,1]$, so as the two following functions $F(w) = L_\lambda^0\left\{T\left[R\sqrt{\frac{1}{n_\lambda^2}+\left(1-\frac{1}{n_\lambda^2}\right)w}\right]\right\}$ and $\overline{\psi}(z') = \psi\left[R\sqrt{1-\left(1-\frac{1}{n_\lambda^2}\right)z'}\right]$, easily leads to the simplified equation

$$\int_{w=0}^{1} F(w)\frac{\cosh\left(\alpha^*\sqrt{w+\frac{z'}{n_\lambda^2}}\right)}{\sqrt{w+\frac{z'}{n_\lambda^2}}}dw = \overline{\psi}(z') \qquad (w,z') \in [0,1]\times[0,1] \tag{21}$$

Eq. (21) is a first kind Fredholm equation of data $\overline{\psi}$ on $[0,1]$ and unknown function $F$ on the same range, with a non symmetric kernel $K(w,z') = \frac{\cosh\left(\alpha^*\sqrt{w+\frac{z'}{n_\lambda^2}}\right)}{\sqrt{w+\frac{z'}{n_\lambda^2}}}$, in the sense where $K(w,z') \neq K(z',w)$. The two symmetric left $K_G$ and right $K_D$ associated kernels are defined by

$$\begin{aligned} K_G(w,z') &= \int_{v=0}^{1} K(v,w)K(v,z')dv \\ K_D(w,z') &= \int_{v=0}^{1} K(w,v)K(z',v)dv \end{aligned} \tag{22}$$

It can be shown that they can be expressed as follows:

$$\begin{aligned} K_G(w,w) &= Chi\left(\frac{2\tau_0\sqrt{n_\lambda^2-1}}{n_\lambda^2}\sqrt{w+n_\lambda^2}\right) - Chi\left(\frac{2\tau_0\sqrt{n_\lambda^2-1}}{n_\lambda^2}\sqrt{w}\right) + \frac{1}{2}\ln\left(\frac{w+n_\lambda^2}{w}\right) \qquad w \neq 0 \\ K_G(w,z') &= Chi\left[\frac{\tau_0\sqrt{n_\lambda^2-1}}{n_\lambda^2}\left(\sqrt{w+n_\lambda^2}+\sqrt{z'+n_\lambda^2}\right)\right] - Chi\left[\frac{\tau_0\sqrt{n_\lambda^2-1}}{n_\lambda^2}\left(\sqrt{w}+\sqrt{z'}\right)\right] \\ &+ Chi\left(\frac{\tau_0\sqrt{n_\lambda^2-1}}{n_\lambda^2}\left|\sqrt{w}-\sqrt{z'}\right|\right) - Chi\left(\frac{\tau_0\sqrt{n_\lambda^2-1}}{n_\lambda^2}\left|\sqrt{w+n_\lambda^2}-\sqrt{z'+n_\lambda^2}\right|\right) \qquad z' \neq w \end{aligned} \tag{23}$$

and

$$K_D(w,w) = n_\lambda^2 \left[ Chi\left( \frac{2\,\tau_0 \sqrt{n_\lambda^2 - 1}}{n_\lambda} \sqrt{w + \frac{1}{n_\lambda^2}} \right) - Chi\left( \frac{2\,\tau_0 \sqrt{n_\lambda^2 - 1}}{n_\lambda} \sqrt{w} \right) + \frac{1}{2} \ln\left( \frac{w + \frac{1}{n_\lambda^2}}{w} \right) \right] \qquad w \neq 0$$

$$K_D(w,z') = n_\lambda^2 \left\{ \begin{array}{l} Chi\left[ \frac{\tau_0 \sqrt{n_\lambda^2 - 1}}{n_\lambda} \left( \sqrt{w + \frac{1}{n_\lambda^2}} + \sqrt{z' + \frac{1}{n_\lambda^2}} \right) \right] - Chi\left[ \frac{\tau_0 \sqrt{n_\lambda^2 - 1}}{n_\lambda} \left( \sqrt{w} + \sqrt{z'} \right) \right] \\ + Chi\left( \frac{\tau_0 \sqrt{n_\lambda^2 - 1}}{n_\lambda} \left| \sqrt{w} - \sqrt{z'} \right| \right) - Chi\left( \frac{\tau_0 \sqrt{n_\lambda^2 - 1}}{n_\lambda} \left| \sqrt{w + \frac{1}{n_\lambda^2}} - \sqrt{z' + \frac{1}{n_\lambda^2}} \right| \right) \end{array} \right\} \qquad z' \neq w \tag{24}$$

where $Chi(x) = \frac{1}{2}\left[ Ei(x) - E_1(x) \right] = \gamma + \ln x + \sum_{k=1}^{+\infty} \frac{x^{2k}}{2k\,(2k)!}$ is the hyperbolic cosine integral [18].

Since *Chi* is a strictly growing function for $x > 0$, $K_D(w,z')$ (respectively $K_G(w,z')$) is a strictly positive function on $]0,1] \times ]0,1]$ and from the previous definition of *Chi*, $K_D(w,z')$ (respectively $K_G(w,z')$) is obviously a continuous function on $]0,1] \times ]0,1]$. Furthermore, a simple analysis shows that if $K_{D,G}$ reaches a minimal value, where $K_{D,G}$ stands indifferently either for $K_D$ or $K_G$, this value is located on the straight line $w = z'$.

Then for the right kernel $K_D$, noting $\Omega = 1.1996787$ the solution of $\tanh \Omega = \frac{1}{\Omega}$, it is possible to prove that if $\tau_0 \leq \frac{n_\lambda \Omega}{\sqrt{n_\lambda^2 - 1}}$, $K_D(w,w)$ is a strictly positive decaying function on $]0,1]$ with $\lim_{\tau_0 \to 0} \left[ \min_{w \in ]0,1]} K_D(w,w) \right] = n_\lambda^2 \ln\left( 1 + \frac{1}{n_\lambda^2} \right)$, and if $\tau_0 > \frac{n_\lambda \Omega}{\sqrt{n_\lambda^2 - 1}}$, $K_D(w,w)$ is a strictly positive function that reaches a minimal value at a given $u_0(\tau_0) \in ]0,1[$ with $\lim_{\tau_0 \to +\infty} u_0(\tau_0) = 0$ and $\lim_{\tau_0 \to +\infty} \left[ \min_{w \in ]0,1]} K_D(w,w) \right] = +\infty$.

The evolution of $K_D(w,z')$ is depicted on figures 11a-d for two absorption coefficients ($1\,m^{-1}$ and $10\,m^{-1}$) and two refractive indices (*1.5* and *4.5*).

In the first case (Figs 11a-b), $\tau_0 \leq \frac{n_\lambda \Omega}{\sqrt{n_\lambda^2 - 1}}$ and the minimal value of $K_D(w,z')$ is $K_D(1,1)$, (*0.860* for $n_\lambda = 1.5$ and *1.032* for $n_\lambda = 4.5$). When the refractive index increases, the values taken by the right kernel significantly increase, but the global behaviour remains unchanged. For $\tau_0 > \frac{n_\lambda \Omega}{\sqrt{n_\lambda^2 - 1}}$ the minimal value of the kernel grows rapidly (Figs 11c-d) when the absorption coefficient increases, since $\min_{w \in ]0,1]} K_D(w,w) = 7.498$ for $n_\lambda = 1.5$ and $\min_{w \in ]0,1]} K_D(w,w) = 12.480$ for $n_\lambda = 4.5$

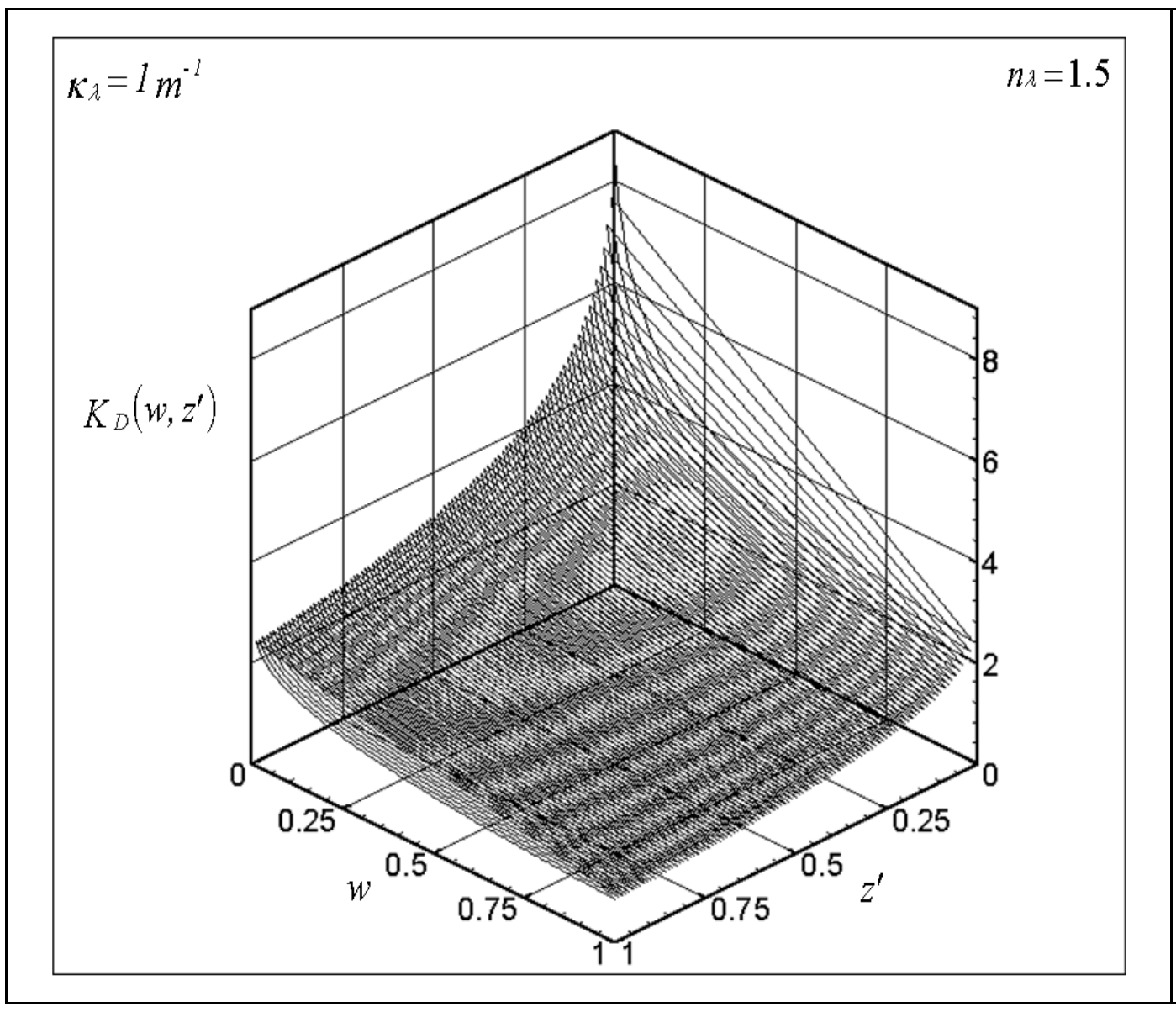

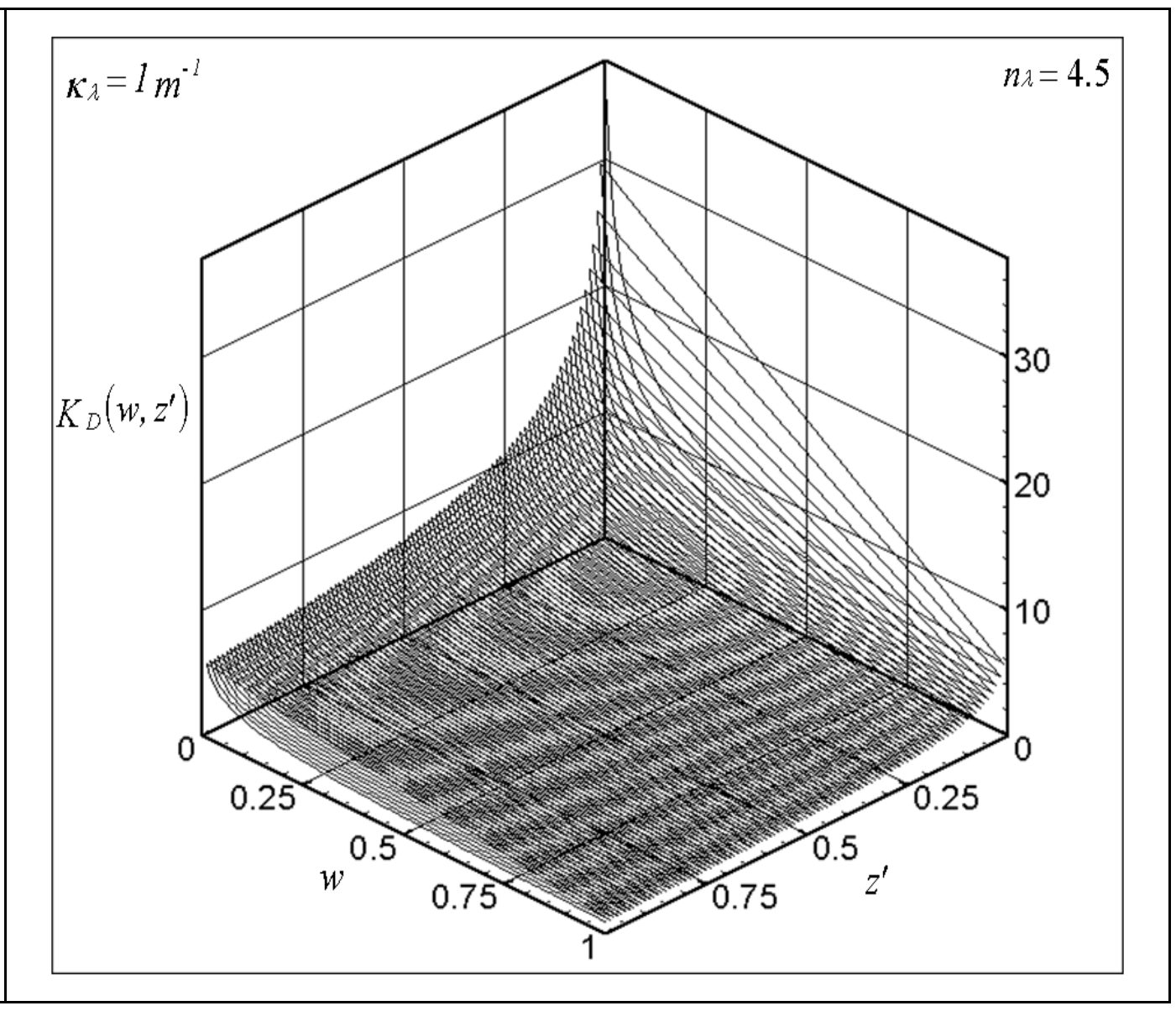

Fig 11a: evolution of kernel $K_D$ for $\tau_0 \leq \frac{n_\lambda \Omega}{\sqrt{n_\lambda^2 - 1}}$

Fig 11b: evolution of kernel $K_D$ for $\tau_0 \leq \frac{n_\lambda \Omega}{\sqrt{n_\lambda^2 - 1}}$

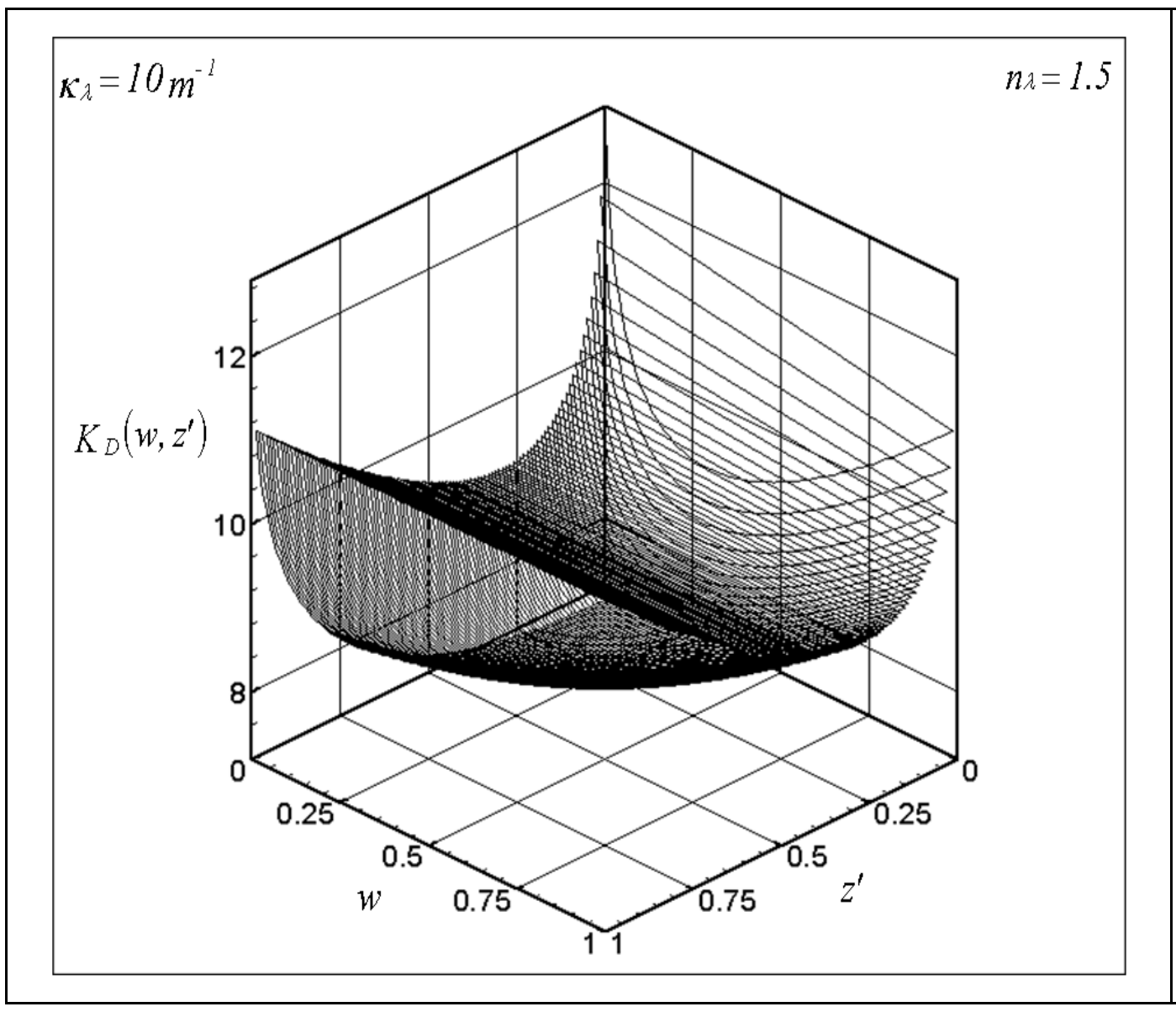

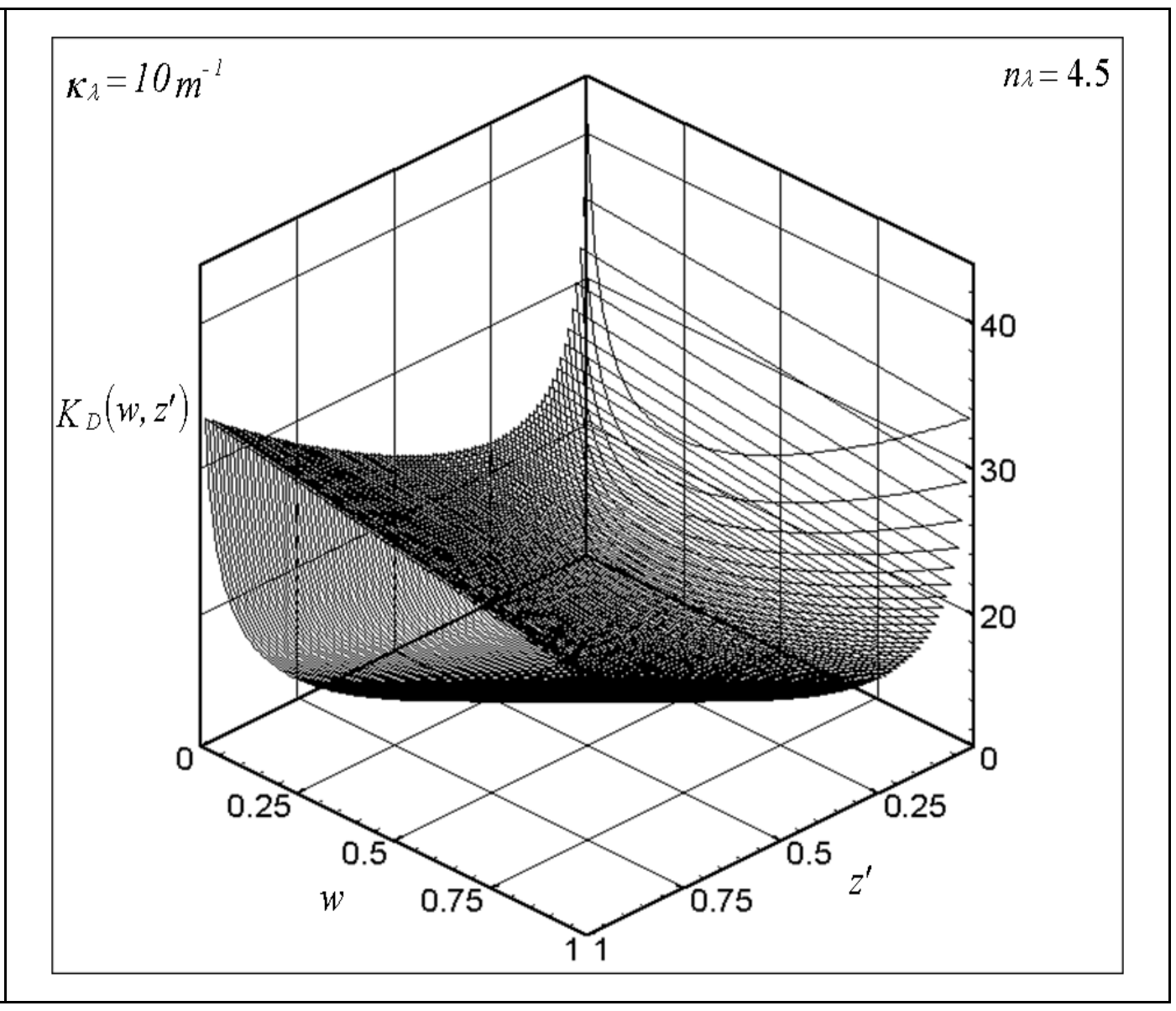

Fig 11c: evolution of kernel $K_D$ for $\tau_0 > \frac{n_\lambda \Omega}{\sqrt{n_\lambda^2 - 1}}$

Fig 11d: evolution of kernel $K_D$ for $\tau_0 > \frac{n_\lambda \Omega}{\sqrt{n_\lambda^2 - 1}}$

The two symmetric kernels are such that $\left\| K_{D,G} \right\|_2^2 = \int_{w=0}^{1} \int_{z'=0}^{1} K_{G,D}^2(w,z')\, dz'\, dw = \Lambda_1^{D,G}(\alpha^*, n_\lambda) < +\infty$, where $\Lambda_1^{D,G}(\alpha^*, n_\lambda)$ (related either to the right or left kernel) are complicated functions in terms of power, logarithm, exponential and product of *Chi* and *Shi* (hyperbolic sine integral) functions of finite value for $\alpha^* \neq 0$ which is the case since $n_\lambda \neq 1$.

Hence $K_{D,G}(w,z')$ are $L_2$-kernels on $[0,1]\times[0,1]$. Obviously, $\overline{\psi}$ is a $L_2$-function on $[0,1]$, so that it guaranties [15] the existence of a solution for Eq. (21) whose formal expression is:

$$F(z') = \underset{n \to +\infty}{lem} \sum_{k=1}^{n} \lambda_k a_k \mu_k(z')$$

$$\Leftrightarrow \lim_{n \to +\infty} \int_{z'=0}^{1} \left[ F(z') - \sum_{k=1}^{n} \lambda_k a_k \mu_k(z') \right]^2 dz' = 0 \qquad (25)$$

where lem commonly stands for the usual limit in mean [19]. The $(\lambda_n)_{n \in |N}$ are the eigenvalues associated to the system $(\mu_n)_{n \in |N}$ of the right kernel eigenfunctions, that is $\mu_n(z') = \lambda_n^2 \int_{w=0}^{1} K_D(w, z') \mu_n(w) dw$, and $(a_n)_{n \in |N}$ are the Fourier coefficients of $\overline{\psi}$ relatively to the system $(\mu_n)_{n \in |N}$, i.e. $\overline{\psi}(z') = \sum_{n=1}^{+\infty} a_n \mu_n(z')$ with $a_n = \int_{z'=0}^{1} \overline{\psi}(z') \mu_n(z') dz'$.

Considering one-normed eigenfunctions, i.e. functions such that $\int_{u=0}^{1} \mu_n^2(u) du = 1$, the eigenvalues associated to the right kernel are such that $\lambda_n^2 \geq \frac{1}{\|K_D\|_2}$.

The exact eigenfunctions and eigenvalues cannot be obtained in an analytical way, so that we shall only determine a numerical approximation of the solution by using the method presented in annex 2.

As an example, the 10 first eigenvalues of the right kernel are listed below, with respect to the numerical quadrature (Gauss quadrature) order *M*, in the case where $R = 0.24\ m$, $\kappa_\lambda = 10\ m^{-1}$ and $n_\lambda = 1.5$, with $Tr\,(K_D) = 8.805$ and $\|K_D\|_2 = 8.743$

| | *M* = 10 | *M* = 100 | *M* = 1000 | *M* = 2000 | *M* = 4000 |
|---|---|---|---|---|---|
| *n* | $\frac{1}{\lambda_n^2}$ | $\frac{1}{\lambda_n^2}$ | $\frac{1}{\lambda_n^2}$ | $\frac{1}{\lambda_n^2}$ | $\frac{1}{\lambda_n^2}$ |
| 1 | 8.740 | 8.743 | 8.743 | 8.743 | 8.743 |
| 2 | $5.167\,10^{-2}$ | $5.821\,10^{-2}$ | $5.822\,10^{-2}$ | $5.822\,10^{-2}$ | $5.822\,10^{-2}$ |
| 3 | $1.295\,10^{-3}$ | $3.840\,10^{-3}$ | $3.876\,10^{-3}$ | $3.876\,10^{-3}$ | $3.876\,10^{-3}$ |
| 4 | $1.520\,10^{-5}$ | $4.393\,10^{-4}$ | $4.806\,10^{-4}$ | $4.807\,10^{-4}$ | $4.807\,10^{-4}$ |
| 5 | $6.152\,10^{-8}$ | $5.589\,10^{-5}$ | $8.250\,10^{-5}$ | $8.267\,10^{-5}$ | $8.269\,10^{-5}$ |
| 6 | $2.870\,10^{-8}$ | $6.438\,10^{-6}$ | $1.727\,10^{-5}$ | $1.750\,10^{-5}$ | $1.754\,10^{-5}$ |
| 7 | $1.754\,10^{-8}$ | $6.426\,10^{-7}$ | $4.030\,10^{-6}$ | $4.259\,10^{-6}$ | $4.312\,10^{-6}$ |
| 8 | $1.344\,10^{-8}$ | $5.548\,10^{-8}$ | $9.668\,10^{-7}$ | $1.124\,10^{-6}$ | $1.178\,10^{-6}$ |
| 9 | $8.018\,10^{-9}$ | $2.315\,10^{-8}$ | $2.261\,10^{-7}$ | $3.043\,10^{-7}$ | $3.347\,10^{-7}$ |
| 10 | $5.692\,10^{-9}$ | $1.865\,10^{-8}$ | $5.063\,10^{-8}$ | $8.107\,10^{-8}$ | $1.028\,10^{-7}$ |

One can see that $\lim_{M \to +\infty} \frac{1}{\lambda_1^2} = \|K_D\|_2$: this result has been obtained for a very large spectrum of various situations, so that one may here postulate that $\frac{1}{\lambda_1^2} = \|K_D\|_2$. Hence, since it is observed that $Tr\, K_D \approx \|K_D\|_2$, one deduces that all the ordered eigenvalues $\frac{1}{\lambda_n^2}$, the first one excepted, quickly tend towards zero with *n*. This main result is strongly amplified when the refractive index increases. Since $\lim_{\kappa_\lambda \to +\infty} \|K_D\|_2 = +\infty$, the first eigen value $\frac{1}{\lambda_1^2}$ has a high magnitude for large absorption coefficients, while the other eigenvalues rapidly tend towards zero. It is then not surprising that the convergence of the eigenvalues is extremely slow. For the considered example, the convergence is only reached for the four first eigenvalues with a 4000 points quadrature. At such an order, because of multiple round-off errors, a Given, Householder or Jacobi reduction completely fails in finding the eigen-elements, and a SVD appears to be the best technique in finding the complete eigen-elements. Nevertheless, due to the extremely slow convergence of the eigenvalues up to a very high quadrature order, the partial sum $f(z) = \sum_{k=1}^{n} \lambda_k a_k \mu_k(z)$ is not correctly evaluated in most cases.

**4.4 Final remarks**

Now, one knows that an exact solution of the problem exists and that no spectral intensities theoretically need to be added to the directional ones to obtain a complete solution on the set $r \in [0, R]$. This means that determining the temperature field inside a cylinder with only directional intensity data, does not belong to the class of "missing data" problem, and that it has a theoretical exact solution. However, from a practical point of view, this exact solution is unreachable for at least two reasons: the experimental data are never perfect, and particularly to this case, the problem is extremely ill-posed because of the intrinsic properties of the kernel governing the Fredholm equation on the set $\left[\frac{R}{n_\lambda}, R\right]$, contrarily to what happens on the set $\left[0, \frac{R}{n_\lambda}\right]$ where a smooth regularisation is most often powerful. Even with sophisticated filtering treatments applied to experimental data, or at the limit case of perfect non noisy data, a numerical approximate solution is very hard to obtain on the $\left[\frac{R}{n_\lambda}, R\right]$, even when using a discretization of the integral equation, followed by a SVD of the discrete operator and zeroing of its too small singular values.

This limitation is essentially critical for media with high refractive indices and of internal temperature fields of complex shape. On the other hand, for usual dense media with indices not too far from *1.5*, and smooth internal temperature fields, a SVD and zeroing works very well in practice with no noisy data. Finally, we can note that a little refinement can be provided to the search of the temperature field on $\left[\frac{R}{n_\lambda}, R\right]$ by adding spectral measurements. Indeed, integrating Eq. (26) on $x \in \left[\frac{R}{n_\lambda}, R\right]$ gives for the 1st order moment:

$$\int_{x=\frac{R}{n_\lambda}}^{R} x \int_{r=\frac{R}{n_\lambda}}^{R} \frac{r\,\kappa_\lambda\, L_\lambda^0[T(r)]}{\sqrt{r^2 - \frac{x^2}{n_\lambda^2}}} \cosh\left(\kappa_\lambda \sqrt{r^2 - \frac{x^2}{n_\lambda^2}}\right) dr\, dx = \int_{x=\frac{R}{n_\lambda}}^{R} x\,\psi(x)\, dx\,, \qquad (26)$$

That is, after a simple integration:

$$\int_{r=\frac{R}{n_\lambda}}^{R} r\left[\cosh\left(\kappa_\lambda \sqrt{r^2 - \frac{R^2}{n_\lambda^4}}\right) - \cosh\left(\kappa_\lambda \sqrt{r^2 - \frac{R^2}{n_\lambda^2}}\right)\right] L_\lambda^0[T(r)]\, dr = \int_{x=\frac{R}{n_\lambda}}^{R} x\,\psi(x)\, dx\,, \qquad (27)$$

which is an integral equation of the form $\int_{r=\frac{R}{n_\lambda}}^{R} K(r)\, L_\lambda^0[T(r)]\, dr = G$, where $G$ is a constant for a given wavelength and $K$ is the kernel depending only on one variable for the same wavelength. Then a discrete form of Eq. (27) can easily be obtained, for several wave lengths where the absorption sensibility is significant.

## 5. CONCLUSION

In this paper, we proved than an exact solution of the problem which consists in determining the temperature field inside a cylinder filled with a dense semi-transparent medium of non unit refractive index, only from directional emerging intensity data, can be theoretically found in the complete domain $[0, R]$ when the useful intensity measurements are perfect non noisy data. A partial solution on the set $\left[0, \frac{R}{n_\lambda}\right]$ can be easily found from a Laplace transform extension applied to the convolution equation governing the problem, and allows to exhibit the generalised "data function *h*", from which the auxiliary function $\psi$ can be determined on the complete set. This latter function is the datum of the 1st kind

Fredholm equation, for which an exact formal solution on $\left[\frac{R}{n_\lambda}, R\right]$ can be proposed. However, even with perfect intensity data, this exact theoretical solution cannot simply be reached on $\left[\frac{R}{n_\lambda}, R\right]$, because of the intrinsic properties of the kernel governing the integral equation to be solved to obtain the corresponding temperature field. Indeed, the ordered eigenvalues of the symmetrised right kernel are such that the first one corresponds to the operator's norm, so that all the other ones quickly tend towards zero since their sum equals the operator's trace minus its norm, quantity extremely close to zero for all absorption coefficients and refractive indices: hence, due to a very slow convergence in the eigenvalues computation, high order quadratures have to be used leading to extremely ill-conditioned discrete operators, for which standard Householder reductions fail in determining the associate eigen functions and solution. That particular behaviour of the kernel and its eigen-elements also explain why a SVD followed by a zeroing of the smallest singular values generally gives poor approximate solutions for high refractive indices and/or absorption coefficients. This means, in other practical words, that in spite of sophisticated filtering and/or regularisation procedures on noisy experimental data, an acceptable approximate solution of the initial problem on the whole set $[0, R]$ cannot be exhibited unless a judicious regularisation of the governing operator itself be applied.

*Acknowledgements:* we greatly appreciated the helpful advice of Pr. Anouar Soufiani

## Annex 1

writing $\displaystyle\int_{s=x_i}^{R} \frac{h'(s)\cos\left(\frac{\kappa_\lambda}{n_\lambda}\sqrt{s^2-x_i^2}\right)}{\sqrt{s^2-x_i^2}}\,ds = \sum_{k=i}^{N-1} I_{ki}$, with

$\displaystyle\frac{I_{ki}}{\Delta r} = \int_{t=0}^{1} \frac{h'(\Delta r\, t + x_k)\cos\left[\frac{\kappa_\lambda}{n_\lambda}\sqrt{(\Delta r\, t + x_k)^2 - x_i^2}\right]}{\sqrt{(\Delta r\, t + x_k)^2 - x_i^2}}\,dt$ for $2 \le i \le N-1$, and noting $H_k(t) = h(\Delta r\, t + x_k)$,

leads to $H'_k(t) = \Delta r\, h'(\Delta r\, t + x_k) = a_k + b_k\, t + c_k\, t^2 + d_k\, t^3$, with a development of $H'_k(t)$ function in Cubic Splines, $H'_k(0) = a_k = \Delta r\, h'_k$, $H'_k(1) = a_k + b_k + c_k + d_k = \Delta r\, h'_{k+1}$, $H''_k(0) = b_k = D_k$, where the $D_k$ are obtained from the linear system:

$$\begin{aligned}
&\sum_{k=1}^{N-1} A_{1,k}\, D_k = 3\,\Delta r\,(h'_2 - h'_1) \\
&\sum_{k=1}^{N-1} A_{i,k}\, D_k = 3\,\Delta r\,(h'_{i+1} - h'_{i-1}) \qquad 2 \le i \le N-1 \\
&\sum_{k=1}^{N-1} A_{N,k}\, D_k = 3\,\Delta r\,(h'_N - h'_{N-1})
\end{aligned} \tag{A1}$$

with $A_{i,k} = \begin{pmatrix} 2 & 1 & 0 & \dots & & & & 0 \\ 1 & 4 & 1 & 0 & \dots & & & 0 \\ 0 & 1 & 4 & 1 & 0 & \dots & & 0 \\ \dots & & & & & & & \\ 0 & \dots & 0 & 1 & 4 & 1 & 0 & 0 \\ \dots & & & & & & & \\ 0 & \dots & & & 0 & 1 & 4 & 1 \\ 0 & 0 & 0 & 0 & 0 & 0 & 1 & 2 \end{pmatrix}$, the coefficients $a_k, b_k, c_k, d_k$ being related to the $D_k$ by

$a_k = \Delta r\, h'_k$, $b_k = D_k$, $c_k = 3\,\Delta r\,(h'_{k+1} - h'_k) - 2\,D_k - D_{k+1}$ and $d_k = 2\,\Delta r\,(h'_k - h'_{k+1}) + D_k + D_{k+1}$. Hence it comes for the previous integrals $I_{ki}$

$$\begin{aligned}
I_{ki} = {} & \frac{n_\lambda}{\kappa_\lambda\,\Delta r}\left\{\left[\tilde{B}_k + \tilde{D}_k\left(v^2 + x_i^2 - \frac{2\,n_\lambda^2}{\kappa_\lambda^2}\right)\right]\sin\left(\frac{\kappa_\lambda\, v}{n_\lambda}\right) + \frac{2\,n_\lambda\,\tilde{D}_k}{\kappa_\lambda}\, v\cos\left(\frac{\kappa_\lambda\, v}{n_\lambda}\right)\right\}_{v=\sqrt{x_k^2 - x_i^2}}^{\sqrt{x_{k+1}^2 - x_i^2}} \\
& + \frac{\alpha}{\Delta r}\int_{u=0}^{1}\left[\tilde{A}_k + \tilde{C}_k\, x_i^2\, ch^2(\alpha\, u + \beta)\right]\cos\left[\frac{\kappa_\lambda}{n_\lambda}\, x_i \sinh(\alpha\, u + \beta)\right] du
\end{aligned} \tag{A2}$$

The different quantities appearing in the $I_{ki}$ integrals are defined by the following relations:

$$\tilde{A}_k = a_k - \frac{b_k\, x_k}{\Delta r} + \frac{c_k\, x_k^2}{\Delta r^2} - \frac{d_k\, x_k^3}{\Delta r^3}, \qquad \tilde{B}_k = \frac{b_k}{\Delta r} - \frac{2\, c_k\, x_k}{\Delta r^2} + \frac{3\, d_k\, x_k^2}{\Delta r^3} = -\frac{d\,\tilde{A}_k}{d\, x_k}, \qquad \tilde{C}_k = \frac{c_k}{\Delta r^2} - \frac{3\, d_k\, x_k}{\Delta r^3} = \frac{1}{2}\frac{d^2\,\tilde{A}_k}{d\, x_k^2},$$

$$\tilde{D}_k = \frac{d_k}{\Delta r^3} = -\frac{1}{6}\frac{d^3\,\tilde{A}_k}{d\, x_k^3},\ \alpha_{ki} = ln\left[\frac{k + \sqrt{k^2 - (i-1)^2}}{k - 1 + \sqrt{(k-1)^2 - (i-1)^2}}\right] \text{ and } \beta_{ki} = ln\left[\frac{k - 1 + \sqrt{(k-1)^2 - (i-1)^2}}{i-1}\right].$$

The integral in $I_{ki}$ being non analytical is computed numerically with a Gauss quadrature. For $i = 1$, the integral $I_{k1}$ simply reduces to:

$$I_{k1} = \frac{n_\lambda}{\kappa_\lambda\, \Delta r}\left\{\left[\tilde{B}_k + \frac{n_\lambda\, \tilde{C}_k}{\kappa_\lambda} + \tilde{D}_k\left(v^2 - \frac{2\, n_\lambda^2}{\kappa_\lambda^2}\right)\right] sin\left(\frac{\kappa_\lambda\, v}{n_\lambda}\right) + \left(\frac{2\, n_\lambda\, \tilde{D}_k}{\kappa_\lambda} - \tilde{C}_k\right) v\, cos\left(\frac{\kappa_\lambda\, v}{n_\lambda}\right)\right\}_{v = x_k}^{x_{k+1}} + \frac{\tilde{A}_k}{\Delta r}\left[Ci\left(\frac{\kappa_\lambda\, x_{k+1}}{n_\lambda}\right) - Ci\left(\frac{\kappa_\lambda\, x_k}{n_\lambda}\right)\right] \qquad \text{(A3)}$$

where $\tilde{A}_1 = 0$, $\tilde{B}_1 = \frac{b_1}{\Delta r}$, $\tilde{C}_1 = \frac{c_1}{\Delta r^2}$, $\tilde{D}_1 = \frac{d_1}{\Delta r^3}$ for $k = 1$.

$Ci$ classically stands for the integral cosine [9], with $Ci(x) = \frac{1}{2}\left[Ei(i\, x) + Ei(-i\, x)\right] = \gamma + ln\, x + \sum_{k=1}^{+\infty} \frac{(-1)^k\, x^{2k}}{2\, k\, (2\, k)!}$, $\gamma$ being the Euler-Mascheroni constant and $Ei$ the exponential integral. The local derivatives of $h$ are then obtained from a centred finite differences scheme, with $h'_1 = 0$, $h'_i = \frac{h_{i+1} - h_{i-1}}{2\, \Delta r}$ for $2 \le i \le N - 1$, and $h'_N = \frac{3\, h_N - 4\, h_{N-1} + h_{N-2}}{2\, \Delta r} = \frac{h_{N-2} - 4\, h_{N-1}}{2\, \Delta r}$. Note here that for the numerical calculation $h'_N$ is forced to have a finite value since one should have $h'_N = -\infty$.

## Annex 2

With the help of a numerical quadrature, the eigen-elements are solution of the discrete equation $\mu_i^n = \lambda_n^2 \sum_{j=1}^{M} \omega_j\, \hat{K}_{ij}^D\, \mu_j^n \Leftrightarrow \mu^n = \lambda_n^2\, \hat{K}^D\, diag\,(\omega)\, \mu^n$, $\hat{K}^D$ being the discrete form of the continuous operator $K_D$, with $\hat{K}_{ij}^D = K_D(w_i, w_j)$. Hence the eigenvalues for the discrete problem are solution of $\left[\hat{K}^D\, diag\,(\omega) - \frac{1}{\lambda_n^2} I\right]\mu^n = 0$, where $0 < \frac{1}{\lambda_n^2} \le \|K_D\|_2$. Since $\|K_D\|_2$ tends quickly towards important values as soon as $\kappa_\lambda$ increases, one writes $\frac{1}{\lambda_n^2} = \alpha_n \left\| K_D \right\|_2$ and $\hat{K}_{ij}^D = \left\| K_D \right\|_2 \tilde{K}_{ij}^D$, and the previous problem is equivalent to $\tilde{K}^D \, diag\,(\omega)\,\mu^n = \alpha_n\,\mu^n$ with $0 < \alpha_n \le 1$, where $\tilde{K}^D$ is a square $M$ x $M$ matrix and $\mu^n$ a $M$ vector.

The global matrix $\tilde{K}^D \, diag\,(\omega)$ of eigenvalues $\alpha_n$ is non symmetric, contrarily to the matrix $diag\,(\omega)\,\tilde{K}^D \, diag\,(\omega)$, whence the initial discrete problem is equivalent to $A\,\mu^n = \alpha_n\, B\,\mu^n$, where $A = diag\,(\omega)\,\tilde{K}^D \, diag\,(\omega)$ is a symmetric matrix and $B = diag\,(\omega)$ is obviously a non singular positive definite matrix, with $B = diag\left(\sqrt{\omega}\right) diag\left(\sqrt{\omega}\right)$. Then the eigenvalues $\alpha_n$ are the ones of the matrix $D = diag\left(\frac{1}{\sqrt{\omega}}\right) A \, diag\left(\frac{1}{\sqrt{\omega}}\right) = diag\left(\sqrt{\omega}\right)\tilde{K}^D \, diag\left(\sqrt{\omega}\right)$, so that the eigenvectors are determined by $\mu^n = diag\left(\frac{1}{\sqrt{\omega}}\right)\nu^n$; the condition for the set $\left(\mu^n\right)_{n \in |N*}$ to be an orthonormal basis in the sense of the $L_2$-scalar product leads to $\int_{u=0}^{1} \mu^n(u)\,\mu^m(u)\,du = \delta_{nm} = \sum_{j=1}^{M} \omega_j\,\mu_j^n\,\mu_j^m = \sum_{j=1}^{M} \nu_j^n\,\nu_j^m$ where $\nu^n$ are the orthonormed eigenvectors (in the usual sense of the scalar product for vectors) of the matrix $D$, so that $\mu^n = diag\left(\frac{1}{\sqrt{\omega}}\right)\nu^n$ are the orthonormed eigenfunctions for the $L_2$-scalar product of the operator $K^D$.

Furthermore, since $\sum_{j=1}^{M} \alpha_j = Tr\left[\tilde{K}^D \, diag\,(\omega)\right] = \sum_{j=1}^{M} \omega_j\,\tilde{K}_{jj}^D$, it obviously comes that $\lim_{M \to +\infty} \sum_{j=1}^{M} \alpha_j = \lim_{M \to +\infty} \sum_{j=1}^{M} \omega_j\,\tilde{K}_{jj}^D = \int_{u=0}^{1} \tilde{K}^D(u,u)\,du = Tr\left(\tilde{K}^D\right)$ and the series $\left(\sum \alpha_n\right)_{n \in |N*}$ is absolutely convergent since $Tr\left(\tilde{K}^D\right) < +\infty$.

Since matrix $D$ is real and symmetric, its eigen-elements can be both obtained thanks to a Jacobi transformation, or it can be reduced to a tridiagonal form with the help of a Given's or Householder's reduction to calculate the eigenvalues and vectors. However matrix $D$ is extremely ill conditioned, with $\lim_{M \to +\infty} (det\,D) = 0$, and even for small $M$'s, only a few eigenvalues is of significant magnitude when the absorption coefficient is moderate. Then Jacobi, Given or Householder reductions fail in finding the eigen elements even at low orders. However, since $D$ is real and symmetric, its eigenvalues equal its singular values obtained by a SVD, the eigenvectors of $D$ being contained in the orthogonal matrix $U$ given by the SVD, such that $D = U \, diag\left(\frac{1}{\lambda_i^2}\right) U^T$.